\documentclass[aps,prb,twocolumn,superscriptaddress,floatfix,showpacs,amsmath,amssymb,nofootinbib,longbibliography]{revtex4-2}

\usepackage{graphicx}
\usepackage{dcolumn,xcolor}
\usepackage{bm}
\usepackage{color}
\usepackage{tabularx}
\usepackage{siunitx}
\usepackage{verbatim}
\usepackage{parskip}
\usepackage{soul}
\usepackage[normalem]{ulem}
\usepackage{array}

\newcommand{\GRS}{GdRu$_2$Si$_2$}
\newcommand{\etal}{\textit{et~al.}}
\newcommand{\tn}{$T_{\rm N}$}
\newcommand{\tone}{$T_{\rm 1}$}
\newcommand{\ttwo}{$T_{\rm 2}$}
\newcommand{\bsat}{$B_{\rm sat}$}

\newcommand{\bskyrin}{$B_{\rm I-II}$}
\newcommand{\bskyrout}{$B_{\rm II-III}$}
\newcommand{\dchiiT}{$\partial (\chi_{\rm i} T)/\partial T$}

\newcommand{\mbfu}{$\mu_\mathrm{B}$/f.u.}
\newcommand{\cpx}{$c_{\rm p}$}
\newcommand{\jmk}{J/(mol\,K)}
\newcommand{\ali}{$\alpha_{\rm i}$}
\newcommand{\alc}{$\alpha_{\rm c}$}

\newcommand{\rk}[1]{\textcolor{black}{#1}}
\newcommand{\luk}[1]{\textcolor{black}{#1}}

\newcolumntype{M}[1]{>{\centering\arraybackslash}m{#1}}

\begin{document}
\title{Uniaxial Pressure Effects, Phase Diagram, and Tricritical Point in the Centrosymmetric Skyrmion Lattice Magnet GdRu$_2$Si$_2$}

\author{L.~Gries}
\affiliation{Kirchhoff Institute of Physics, Heidelberg University, INF 227, D-69120 Heidelberg, Germany}

\author{T.~Kleinbeck}
\affiliation{Kirchhoff Institute of Physics, Heidelberg University, INF 227, D-69120 Heidelberg, Germany}

\author{D.A.~Mayoh}
\affiliation{Department of Physics, University of Warwick, Coventry, West Midlands
CV4 7AL, United Kingdom}

\author{G.D.A.~Wood}
\affiliation{Department of Physics, University of Warwick, Coventry, West Midlands
CV4 7AL, United Kingdom}

\author{G.~Balakrishnan}
\affiliation{Department of Physics, University of Warwick, Coventry, West Midlands
CV4 7AL, United Kingdom}
\author{R.~Klingeler}
\email{klingeler@kip.uni-heidelberg.de}\affiliation{Kirchhoff Institute of Physics, Heidelberg University, INF 227, D-69120 Heidelberg, Germany}
\date{\today}

\begin{abstract}

The magnetic phase diagram, magnetoelastic coupling, and uniaxial pressure effects of centrosymmetric magnetic skyrmion-hosting \GRS\ are investigated by means of high-resolution capacitance dilatometry in fields up to 15\,T supported by specific heat and magnetisation studies. In addition to the previously reported phases in the $H$-$T$ phase diagram, we observe a third antiferromagnetic phase in zero magnetic field. We present the magnetic phase diagram and find two unreported phases, one of which features a comparably giant uniaxial pressure dependence. Our dilatometric measurements show magnetoelastic effects associated with the various magnetic ordering phenomena. We determine the uniaxial pressure dependencies of the various phases, in particular of the skyrmion lattice phase which is enhanced at higher fields and temperatures and also widens at a rate of 0.07~T/GPa when uniaxial pressure is applied along the $c$ axis. The relevance of fluctuations is further highlighted by the presence of tricritical point \luk{indicated by our thermodynamic data} at the phase boundary separating two double-\textit{Q} magnetic configurations between which the skyrmion pocket phase evolves upon further cooling. 

\end{abstract}

\maketitle

\section{Introduction}

Magnetic skyrmions are a class of topologically protected non-collinear spin structures which exhibit a variety of particle-like properties~\cite{Skyrme.1961, Skyrme.1962, Bogdanov.1989, Nagaosa.2013, Buttner.2015, Iwasaki.2013}. The nanometric scale of magnetic skyrmions and emergence of interesting phenomena like the topological Hall effect \cite{Lee.2009, Yin.2015, Leroux.2018} and non-linear tunneling magneto-resistance \cite{Hanneken.2015} lead to the proposition of novel skyrmion based technological applications including neuromorphic computation systems \cite{Prychynenko.2018, Song.2020, Yokouchi.2022}, non-volatile memory \cite{Luo.2021, Koshibae.2015, Yu.2017} and logical gates~\cite{Luo.2018, Zhang.2015b, Sisodia.2022}. Since the initial experimental finding of magnetic skyrmions in MnSi in 2009 by Mühlbauer \etal~\cite{Muhlbauer.2009} a plethora of systems hosting a skyrmion lattice have been discovered, e.g., Cu$_2$OSeO$_3$~\cite{Seki.2012}, GaV$_4$S$_8$~\cite{Kezsmarki.2015} and thin film Fe$_{\rm 0.5}$Co$_{\rm 0.5}$Si~\cite{Yu.2010}. In these non-centrosymmetric systems the Dzyaloshinskii–Moriya interaction is an integral part of the skyrmion formation process. However this is not the case in selected centrosymmetric systems such as Gd$_2$PdSi$_3$~\cite{Kurumaji.2019}, Gd$_3$Ru$_4$Al$_{12}$~\cite{Hirschberger.2019}, EuAl$_4$~\cite{Takagi.2022} and \GRS~\cite{Khanh.2020, Yasui.2020} where various stabilisation mechanisms from geometric frustration \cite{Leonov.2015, Okubo.2012} to multiple spin interactions \cite{Hayami.2017} have been proposed. These centrosymmetric system are interesting for the aforementioned technological applications due to their small skyrmion diameter on the order of $\sim 2$\,nm \cite{Yasui.2020, Hirschberger.2019, Luo.2021}.

\GRS\ crystallises in centrosymmetric tetragonal structure in the spacegroup I4/mmm \cite{Hiebl.1983, Felner.1984, Slaski.1984}. It consists of square layers of Gd$^{3+}$ ions ($S = 7/2$, $L = 0$) and Ru$_2$Si$_2$ layers stacked alternatingly along the $c$ axis. In zero magnetic field, two distinct magnetic phases have been reported: phase IV (magnetic phases are labeled in accordance with \cite{Wood.2023}) which evolves at \tn~$\simeq 46$\,K and phase I below $T_{\rm t} \simeq 40$\,K \cite{Garnier.1995, Khanh.2020}. Neutron diffraction experiments reveal that phase IV assumes a sinusoidal or helical spin structure \cite{Paddison.2024}, while phase I, the ground state, forms a double-\textit{Q} constant moment structure \cite{Wood.2023}. Applying magnetic fields parallel to the $c$ axis induces a first order phase transition at about $2$\,T below $20$\,K \cite{Garnier.1995}. It features a double-\textit{Q} magnetic structure consisting of the superposition of two orthogonal helices creating a skyrmion lattice (SKL) in this field-induced phase II \cite{Khanh.2020, Yasui.2020}. By further increasing $B||c$, the SKL phase is replaced by a double-\textit{Q} magnetic structure in phase III above $2.4$\,T while full polarisation of the spin structure is achieved for $B||c \gtrsim 10$\,T~\cite{Khanh.2020}. The origin of the observed magnetic structures has been mainly attributed to a combination of RKKY and multiple spin interactions \cite{Khanh.2022, Yasui.2020, Bouaziz.2022, Eremeev.2023}.

Here we report magneto-elastic coupling and the initial uniaxial pressure dependencies on the various phases in \GRS\ as well as the discovery of two new magnetic phases VI and VII. We achieve this by means of high-resolution capacitance dilatometry studies to determine thermal expansion and magnetostriction in fields up to $B||c =14$\,T which are supported by magnetisation and specific heat measurements. In addition to further completing the magnetic phase diagram, we report the thermodynamic properties at the phase boundaries including evidence of a tricritical point and widening of the SKL phase upon application of uniaxial pressure $p||c$. Specifically, the onset temperature of the SKL phase, at $B||c = 2.1$\,T, increases by $\partial T_{\rm II-III}/\partial p_{\rm c} = 8.5(1.2)$\,K/GPa and the field range where it is present at 2~K widens as $\Delta B_{\rm skyr}/p_{\rm c} \approx 0.07$\,T/GPa.

\section{Experimental Methods}

All measurements were performed on two oriented cuboid-shaped single crystals of \GRS . The cuboids were cut from the same single crystalline boule which has been grown by the floating-zone method as described in Ref. \cite{Wood.2023}. The sample dimensions and orientations were $0.590~(||a) \times 0.801~(||b) \times 1.182~(||c)$\, mm$^3$ (crystal 1) and $0.761~(||[110]) \times 1.400~(||[-110]) \times 3.112~(||c)$\, mm$^3$ (crystal 2). Magnetisation measurements were performed using the vibrating sample magnetometer option (VSM) of Quantum Design's Physical Properties Measurement System (PPMS-14) and Quantum Design's Magnetic Properties Measurement System (MPMS3) in the temperature range of $2$\,K to $300$\,K and in fields up to $14$\,T and $7$\,T respectively. Specific heat data were obtained in the range of $1.8$\,K to $300$\,K \rk{and with high resolution on the smaller crystal 1 used for dilatometry in the range of 1.8\,K to 50\,K} by means of a relaxation method using the PPMS. In order to account for differences in the absolute values due to mass and background errors, the latter data were scaled to the full temperature-range data at 50\,K (see Fig.~S1 in the SM~\cite{supplement}). High-resolution dilatometric measurements were performed in two different setups both using a three-terminal capacitance dilatometer from Küchler Innovative Measurement Technologies \cite{Kuchler.2012, Kuchler.2017}. The first setup is home-built and placed in a Variable Temperature Insert (VTI) of an Oxford Instrument magnet system \cite{Werner.2017}. The second setup is an insert provided from Küchler Innovative Measurement Technologies for the PPMS including an option to rotate the sample up to $90^{\circ}$. The linear thermal expansion coefficients $\alpha_i = 1/L_{\rm i}\times dL_{\rm i}(T)/dT$ are derived from the relative length changes. Furthermore, magnetostriction, \textit{i.e.}, the field induced relative length changes $dL_{\rm i}(B)/L_{\rm i}$ are obtained for fields up to $15$\,T at various temperatures up to $150$\,K and the magnetostriction coefficients $\lambda_i = 1/L_{\rm i}\times dL_{\rm i}(B)/dB$ are derived. For all dilatometric measurements the field is aligned parallel to the measured axis. 

\section{Results}

\subsection{Magnetic order and magnetoelastic coupling in zero magnetic field}\label{zero}

The static magnetic susceptibility $\chi_{\rm i}$ along the different crystallographic directions $i$ follows a Curie-Weiss-like behaviour down to $\sim 85$\,K as shown in Fig.~\ref{fig:chi-T} as well as in Fig.~S2 in the Supplemental Material (SM)~\cite{supplement}. The high-temperature behaviour evidences an isotropic $g$-factor as expected for Gd$^{3+}$ systems. Upon further cooling, non-linear behaviour of $\chi^{-1}$ below about 80~K and appearance of small anisotropy indicates the evolution of short range magnetic order. Pronounced peaks in $\chi_{\rm i}$ at \tn\ $= 45.7(5)$\,K signal the onset of the long range antiferromagnetic order. Below \tn\ two additional anomalies evidence additional magnetic phase transitions at $T_1 = 44.6(5)$\,K and $T_2 = 39.0(5)$\,K, respectively. All observed phase transitions are of a continuous nature and are clearly visible as distinct jumps in Fisher's specific heat \dchiiT\ (see the inset of Fig.~\ref{fig:chi-T}). Note, that anomalies at \tn\ and \ttwo\ have been observed before while the phase transition at \tone\ has not been reported yet~\cite{Khanh.2022, Yasui.2020,Garnier.1995}. Fitting the averaged static magnetic susceptibility $\chi = (2\chi_{\rm a} + \chi_{\rm c})/3$ between $200$ and $300$\,K by means of an extended Curie-Weiss law $C/(T+\Theta)+\chi_0$ yields the effective moment $\mu_{\rm eff} = 8.0(2) \,\mu_{\rm B}$/f.u. which agrees well with previous reports~\cite{Samanta.2008} and matches the expected value for free Gd$^{3+}$ moments of $7.94\, \mu_{\rm B}$/f.u.. The obtained Weiss temperature $\Theta = 42(1)$\,K indicates predominant ferromagnetic interaction in \GRS. 

\begin{figure}[tb]
    \centering
    \includegraphics[width=1\columnwidth,clip]{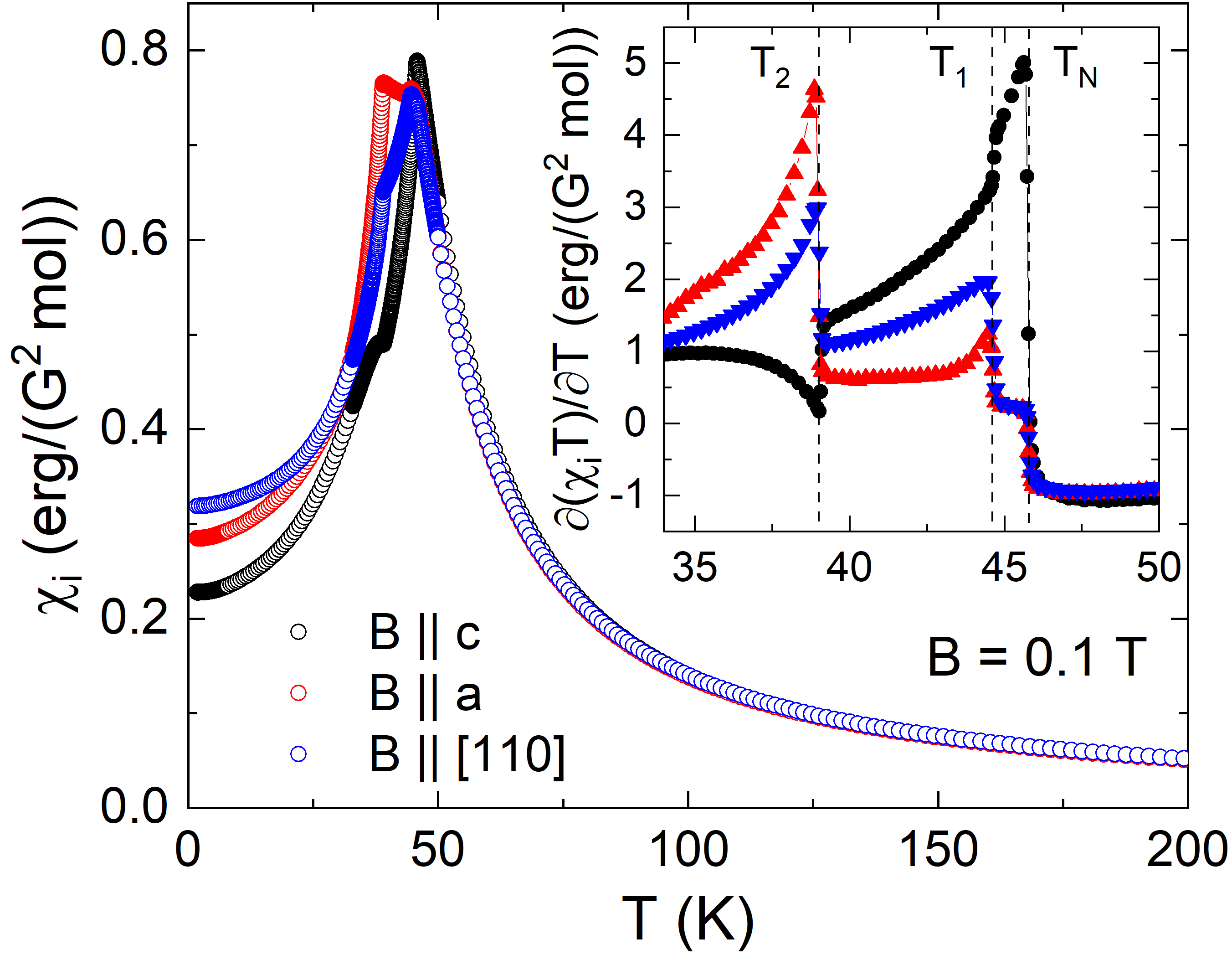}
    \caption{Temperature dependence of the static magnetic susceptibility $\chi = M/B$ measured at $B = 0.1$\,T applied along the crystallographic $i$ axis ($i = c,a,[110]$). The inset shows Fisher's specific heat $\partial (\chi_iT)/\partial T$. Dashed lines mark the anomalies at \tn , \tone , and \ttwo\ (see the text).}
    \label{fig:chi-T}
\end{figure}

\begin{figure}[htb]
    \centering
    \includegraphics[width=1\columnwidth,clip]{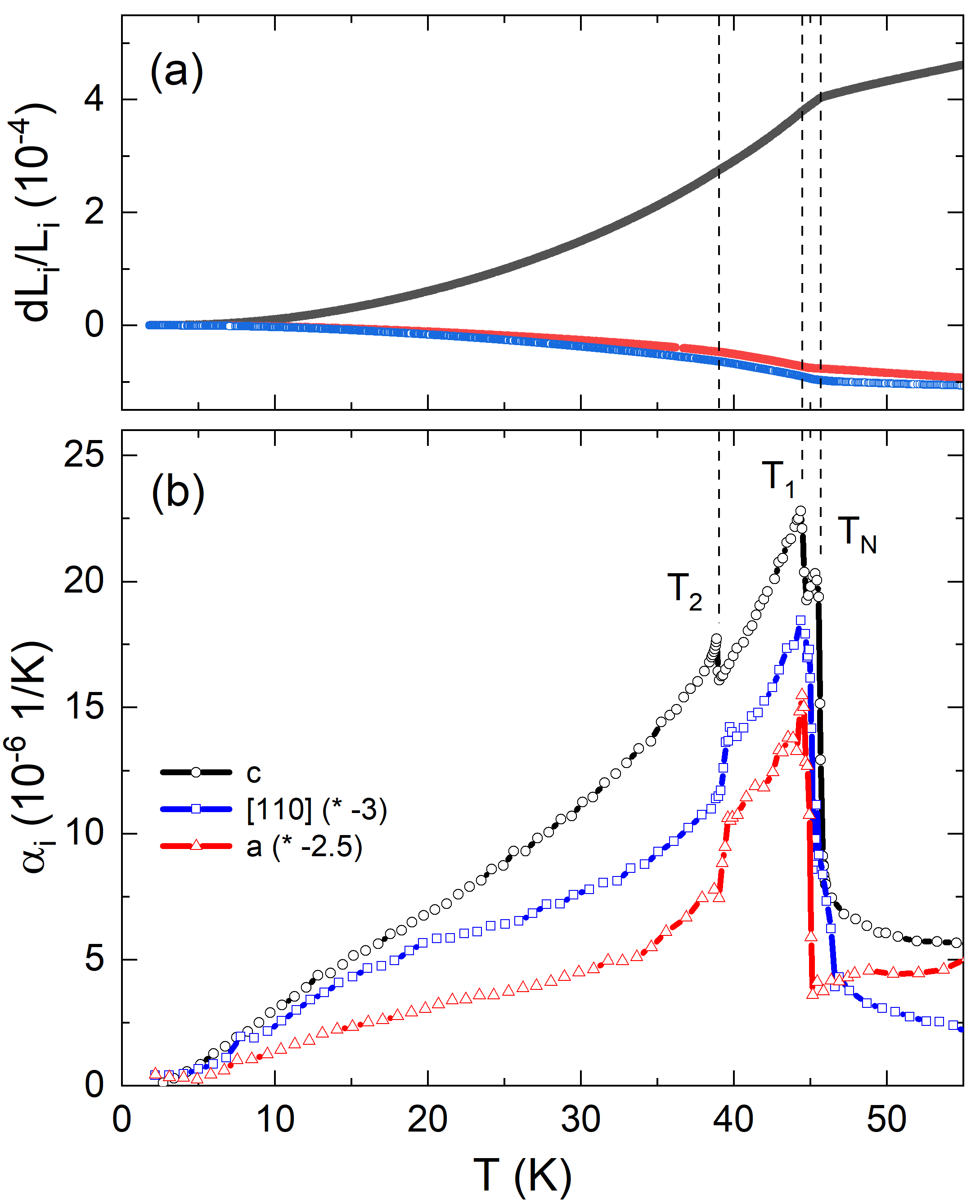}
    \caption{(a) Relative length changes $dL_i/L_i$ along the crystallographic $c$, $[110]$ and $a$ axes, and (b) corresponding thermal expansion coefficients \ali . For better visualisation, data for $\alpha_{\rm [110]}$ and $\alpha_{\rm a}$ have been multiplied by -3 and -2.5, respectively. Vertical dashed lines indicate the transition temperatures \tn , \tone , and \ttwo .}
    \label{fig:TE-0T}
\end{figure}

The evolution of antiferromagnetic order in zero magnetic field is accompanied by pronounced anomalies in the relative length changes $dL_i/L_i$ and in the thermal expansion coefficients \ali\ as can be seen in Fig.~\ref{fig:TE-0T}. Monotonous shrinking of the $c$ axis upon cooling, which we observe in the whole temperature regime under study up to 200~K, is superimposed by a clear kink at \tn. The corresponding thermal expansion coefficient \alc\ shows three positive $\lambda$-like anomalies at \tn, \tone\ and \ttwo. This observation agrees to a previous study where however only one anomaly, at \tn , was observed in \alc ~\cite{Prokleska.2006}. Our data imply significant magnetoelastic coupling since the onset and changes of magnetic order are accompanied by distinct lattice changes. Similarly, there are also clear anomalies in $dL_i/L_i$ at \tn\ for the $a$ and $[110]$ axis. Both axes show monotonous expansion upon cooling below 60~K. Again, three jump-like discontinuities can be seen in the thermal expansion coefficient $\alpha_{[110]}$ indicating the distinct phase boundaries. The associated jumps upon cooling are negative at \tn\ and \tone\ but the one at \ttwo\ is positive (see Fig.~\ref{fig:TE-0T} but note the negative scaling factor). For the $a$ axis only two jumps are observed in $\alpha_{\rm a}$: a negative one at \tn\ and a positive one at \ttwo. This indicates that the dependence of \tone\ on uniaxial pressure along the $a$ axis is small so that the anomaly is not resolved in our measurement. 

The Ehrenfest relation

\begin{equation} \label{eq:Ehrenfest_2} 
\left.\frac{{\partial}T^*}{{\partial}p_i}\right|_B = T^*V_{\mathrm{m}}\frac{\Delta\alpha_i}{\Delta c_{\mathrm{p}}}
\end{equation}

links the anomalies $\Delta\alpha_i$ and $\Delta c_{\rm p}$ of a continuous phase transition to the uniaxial pressure dependence of the transition temperature $T^*$. Hence, the signs of thermal expansion anomalies imply the respective signs of the corresponding initial uniaxial pressure dependencies. The data in Fig.~\ref{fig:TE-0T}b hence signal the increase of \tn, \tone\ and \ttwo , respectively, upon application of uniaxial pressure along the $c$ axis, \textit{i.e.}, $\partial T_{\rm N/1/2} / \partial p_{\rm c} > 0$. Likewise, we read off that uniaxial pressure applied along the $a$ and $[110]$ axis, respectively, will yield a decrease of \tn. This result is in agreement with theoretical predictions by Bouaziz~\etal~\cite{Bouaziz.2022} which propose decrease of \tn\ when compressing the lattice along the $a$ direction. Notably, our data show that the effect of in-plane uniaxial pressure on \ttwo\ is opposite to the effect on \tn\ as both $p_{[110]}$ and $p_{\rm a}$ yield an increase of \ttwo\ (\textit{i.e.}, $\partial T_{\rm 2} / \partial p_{\rm a/[110]} > 0$).

\begin{figure}[tb]
    \centering
    \includegraphics[width=1\columnwidth,clip]{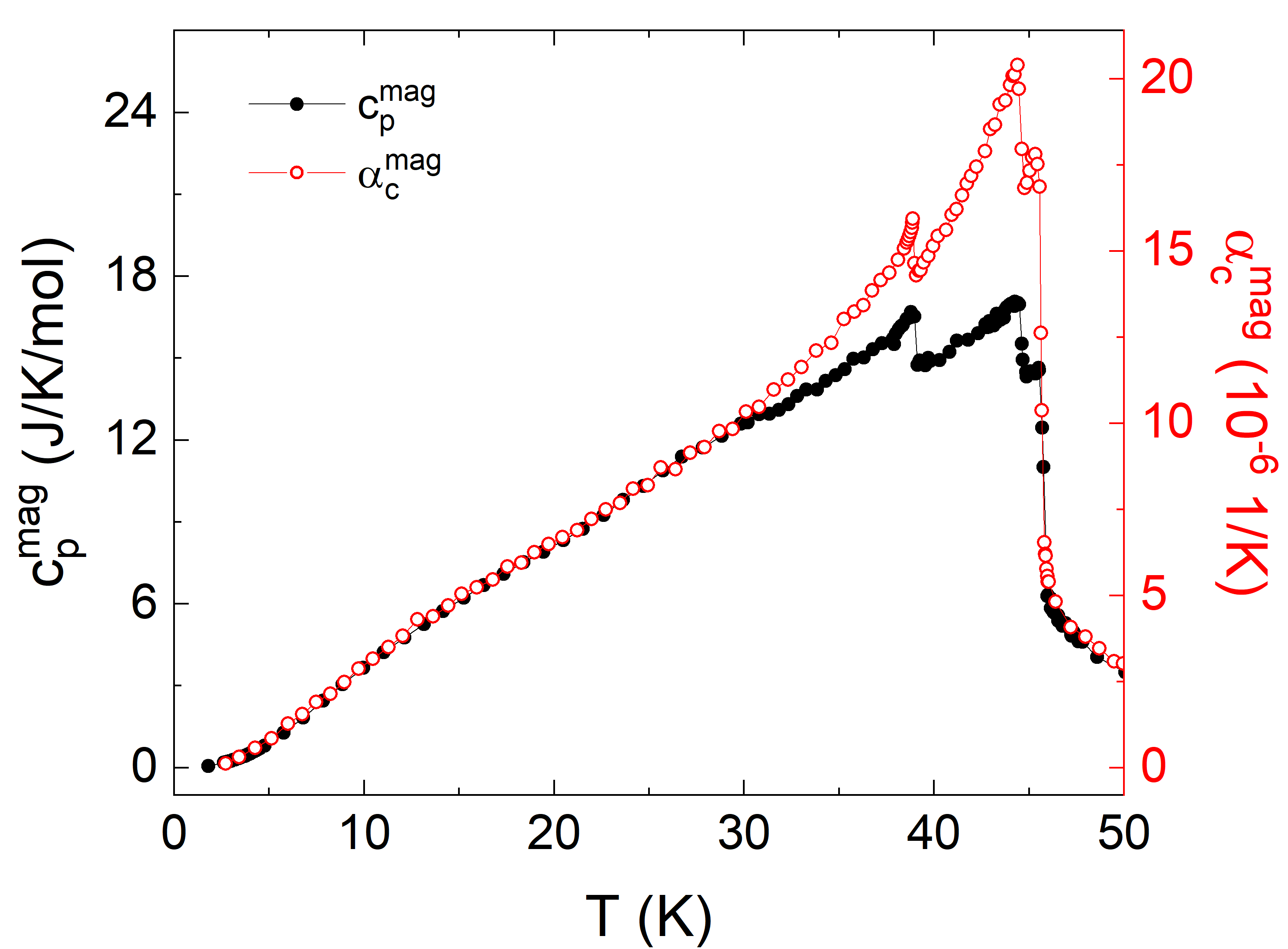}
    \caption{Magnetic specific heat $c_{\rm p}^{\rm mag}$ (filled black markers; left ordinate) and  magnetic contribution to the thermal expansion coefficient $\alpha_{\rm c}^{\rm mag}$ (open red markers; right ordinate). }
    \label{fig:GS}
\end{figure}

To further investigate the successive ordering phenomena, the magnetic Grüneisen parameter $\gamma_{\rm i} = \alpha_{\rm i}^{\rm mag}/c_{\rm p}^{\rm mag}$ is evaluated which compares the magnetic contribution to the thermal expansion coefficient $\alpha_{\rm i}^{\rm mag}$ and the magnetic heat capacity $c_{\rm p}^{\rm mag}.$~\cite{Gegenwart2016,Hoffmann2021} The magnetic heat capacity is determined by subtracting the phononic contribution to the heat capacity $c_{\rm p}^{\rm phon}$ from the experimental data, \textit{i.e.}, $c_{\rm p}^{\rm mag} = c_{\rm p} - c_{\rm p}^{\rm phon}$. In order to estimate the phononic contribution an Einstein-Debye model with one Einstein and one Debye mode is fitted to $c_{\rm p}$ \luk{between $80$\,K and $300$\,K (see Fig.~S3 in the SM)}. The model yields $\Theta_{\rm E} = 598$\,K and $\Theta_{\rm D} = 283$\,K. \luk{Calculating the resulting magnetic entropy by integrating $(c_{\rm p} - c_{\rm p}^{\rm ph}) / T$ yields $S_{\rm mag} = 16.6$\,J/(mol\,K), which is in good agreement with the expected magnetic entropy changes of a Gd$^{3+}$ system $S_{\rm mag}^{\rm theo} = {\rm R}\times \ln (8) = 17.3$\,J/(mol\,K) and confirms the reliability of the thus obtained background. The parameters $\Theta_{\rm i}$} were then used to approximate the phononic contribution to the thermal expansion coefficients \luk{with the prefactors as free fitting variables}\footnote{For a detailed description of the general procedure see Ref.~\cite{Spachmann2022}.}. Figure~\ref{fig:GS} shows the resulting magnetic contributions of $c_{\rm p}$ and $\alpha_{\rm c}$ scaled such that they overlap at low temperatures. Our results imply that the magnetic Grünseisen ratio is constant up to $\sim 30$\,K with the Grüneisen parameter $\gamma_{\rm c} = 7.9(2)\times 10^{-7}$\,mol/J. We conclude the presence of a single dominant energy scale $\epsilon$ for this temperature regime which uniaxial pressure dependence can be derived by using the Grüneisen relation~\cite{Klingeler2006,Gegenwart2016}

\begin{equation}  
 \frac{\partial \ln(\epsilon)}{\partial p_i} = V_{\rm m}\frac{\alpha_{\rm i}}{c_{\rm p}}.\label{gruen}
\end{equation}

Using the molecular volume $V_{\rm m} = 5.02(1) \cdot 10^{-5}$\,m$^3$/mol~\cite{Felner.1984}, applying Eq.~\ref{gruen} yields $\partial ln(\epsilon) / \partial p_{\rm c} = 4.0(1)$~\%/GPa. In contrast, Grüneisen scaling by a single parameter fails for temperatures above $30$\,K \luk{up to \tn}. This implies that neither in phase IV ($T_{\rm 2}\leq T\leq T_{\rm 1}$) nor in phase VI ($T_{\rm 1}\leq T\leq T_{\rm N}$) magnetic order is driven by a single energy scale but there are competing degrees of freedom in both phases. In addition, the experimental observation that Grüneisen scaling fails at $T\gtrsim 30$\,K, \textit{i.e.}, well below \ttwo , implies the presence of a competing energy scale in this temperature regime of the low-temperature phase I, too~\cite{Werner2019}. 

\begin{table}[tb]
\centering
\caption{Jumps (see text as well as Fig.~S4) in the thermal expansion coefficient \alc\ and the heat capacity \cpx\ at $T_{\rm j}$ ($j = N,1,2$) as well as the resulting uniaxial pressure dependencies calculated using the Ehrenfest relation (Eq.~\ref{eq:Ehrenfest_2}).}
\begin{tabular}{c c c c }
	  \hline \hline
		  & $\Delta \alpha_{\rm c}$ [$10^{-6}$\,1/K] & $\Delta c_{\rm p}$ [J/K/mol]& $\partial T_{\rm j} / \partial p_{\rm c}$  [K/GPa]   \\
		\hline
	    \tn\ &   13.2(1.2)  &  9.2(1.2)       &  3.3(5)	          \\
        \tone\ &  3.3(3)   &  2.8(4)      &  2.6(4)	         \\
		\ttwo\ &  1.2(2)   &  1.9(3)       &  1.2(3)	        \\
 \hline \hline
\end{tabular}
\label{tab:ehrenfest}
\end{table}

Comparing the jumps in the thermal expansion coefficient $\Delta \alpha_{\rm i}$ and heat capacity $\Delta c_{\rm p}$ at the phase transitions enables us to quantify the uniaxial pressure dependence of the respective transition via the Ehrenfest relation (Eq.~\ref{eq:Ehrenfest_2}). For a $\lambda$-shaped anomaly, the respective jumps are superimposed by fluctuations so that the height of the jumps at the phase transition are determined by fitting lines to the data in a regime below and above the phase transition temperatures respectively \cite{Kuchler.2005} (see Fig.~S4 in the SM). The resulting jumps in \alc\ and \cpx\ as well as the calculated uniaxial pressure dependencies for uniaxial pressure along the $c$ axis are listed in Table~\ref{tab:ehrenfest}.

\subsection{Effect of magnetic fields $B \parallel c$ and magnetic phase diagram}

\begin{figure}[tb]
    \centering
    \includegraphics[width=1\columnwidth,clip]{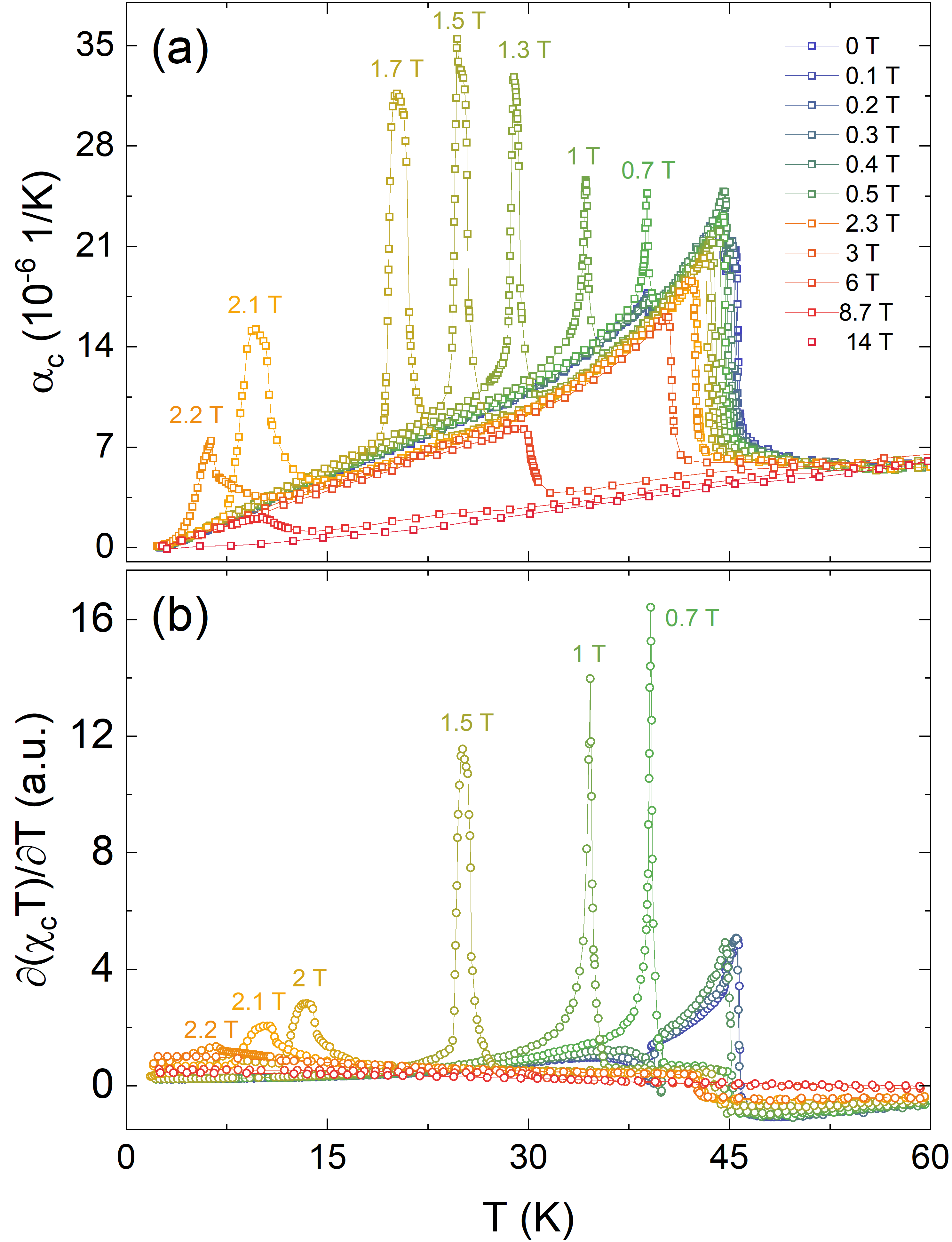}
    \caption{(a) Linear thermal expansion coefficient \alc\ for various magnetic fields $B||c$ up to 14\,T. (b) Fisher's specific heat $\partial (\chi_{\rm c} T)/\partial T$ for various magnetic fields up to $8.7$\,T. For a more detailed view of each individual measurement see Fig.~S6 and Fig.~S8 in the SM. }
    \label{fig:TE-field}
\end{figure}

\begin{figure*}[htb]
    \centering
    \includegraphics[width=2\columnwidth,clip]{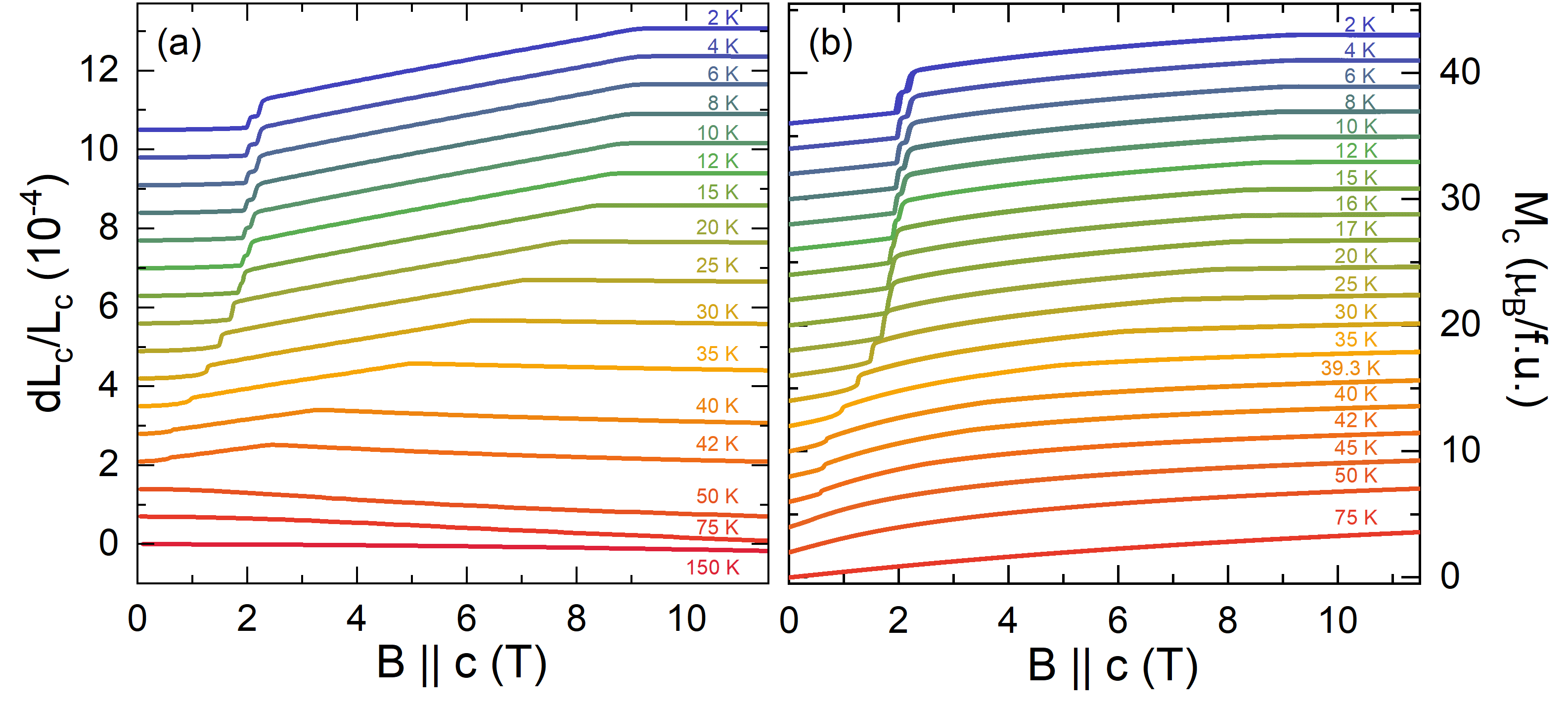}
    \caption{(a) Magnetostriciton $dL_{\rm c}(B)/L_{\rm c}$ and (b) isothermal magnetisation $M_{\rm c}$ at various temperatures as a function of the magnetic field $B \parallel c$. For a detailed plot of the low-field region see Fig.~S9 and Fig.~S11 in the SM.}
    \label{fig:MS-MB-all}
\end{figure*}

\begin{figure*}[htb]
    \centering
    \includegraphics[width=1.8\columnwidth,clip]{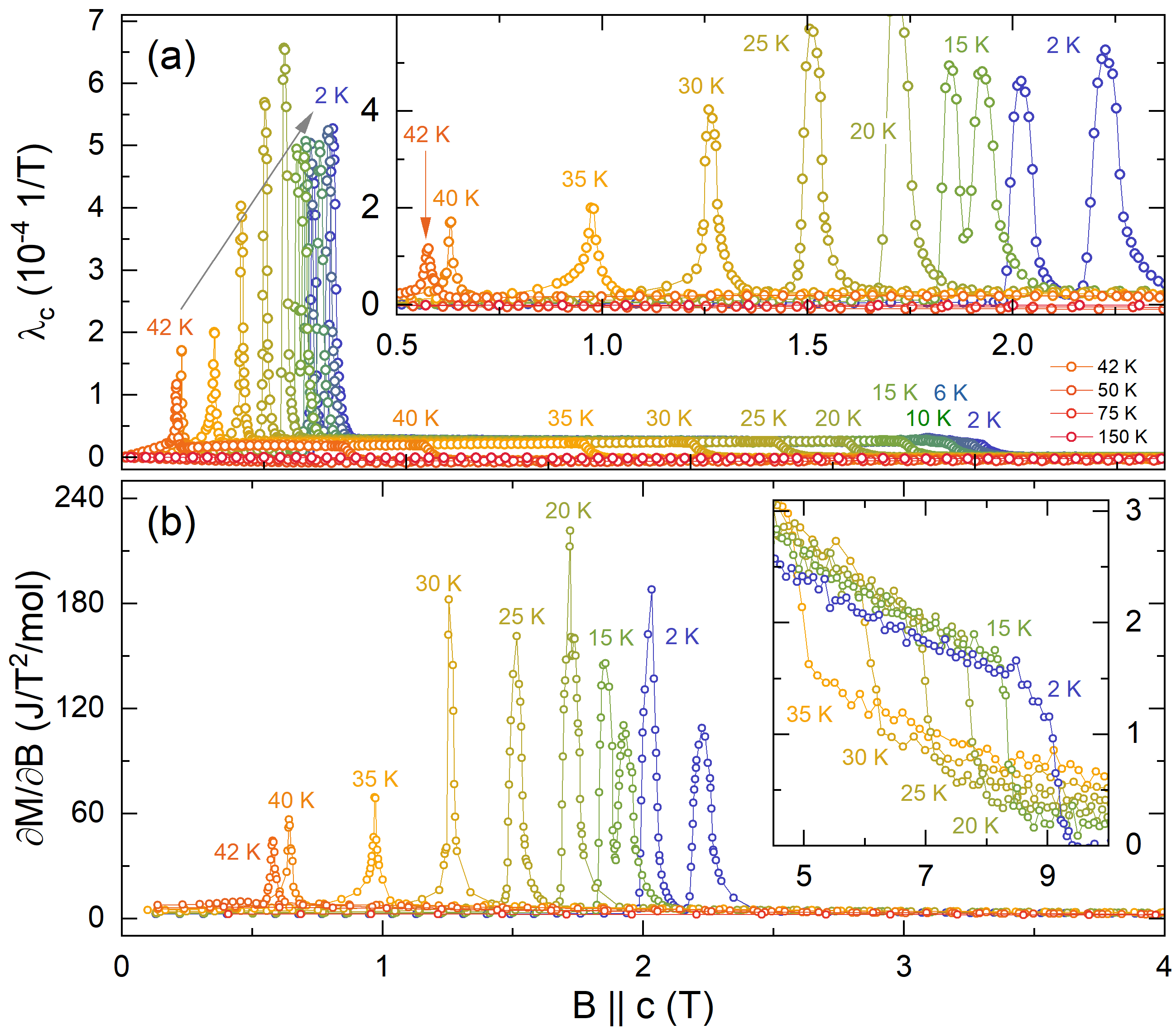}
    \caption{(a) Magnetic field dependence of the magnetostriction coefficient $\lambda_{\rm c}$ for $B \parallel c$ at temperatures between $2$\,K and $150$\,K. The inset shows the region between $0.5$\,T and $2.3$\,T in more detail. (b) Magnetic susceptibility $\partial M_{\rm c}/ \partial B$ at various temperatures for $B \parallel c$. The inset shows the behaviour at the fully polarised state in more detail. \rk{See Figs.~S8 and S10 in the SM~\cite{supplement} for separate plots of the data.}}
    \label{fig:lambda_dMdB}
\end{figure*}

\begin{figure}[htb]

    \centering
    \includegraphics[width=1\columnwidth,clip]{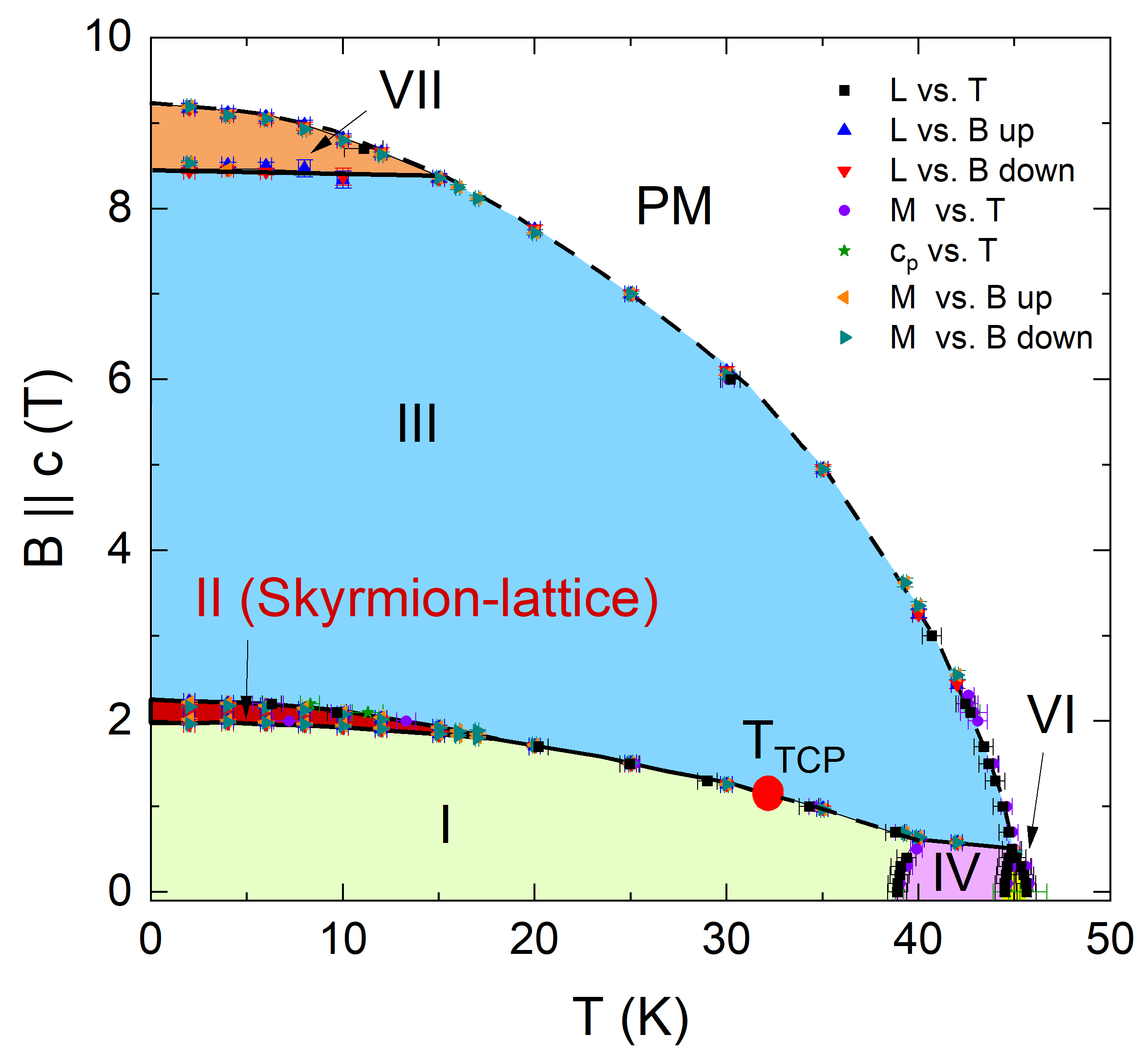}
    \caption{Magnetic phase diagram of \GRS\ for $B||c$ axis constructed from magnetisation $M(T,B)$, dilatometry $L(T,B)$ and specific heat $c_{\rm p}(T,B)$ data. 
    Lines are guides to the eye. Solid/dashed lines represent first/second order phase boundaries; $T_{\rm TCP}$ approximates the position of the tricritical point. Phase II (red) marks the skyrmion lattice phase, PM (white) the paramagnetic phase, phase I (green) and III (blue) are double-\textit{Q}-states. The spin configuration in phase IV (purple) has not been resolved yet; phases VI (yellow) and VII (orange) have not been reported before.}
    \label{fig:phasediagram}

\end{figure}

\begin{figure}[h!]
    \centering
    \includegraphics[width=1\columnwidth,clip]{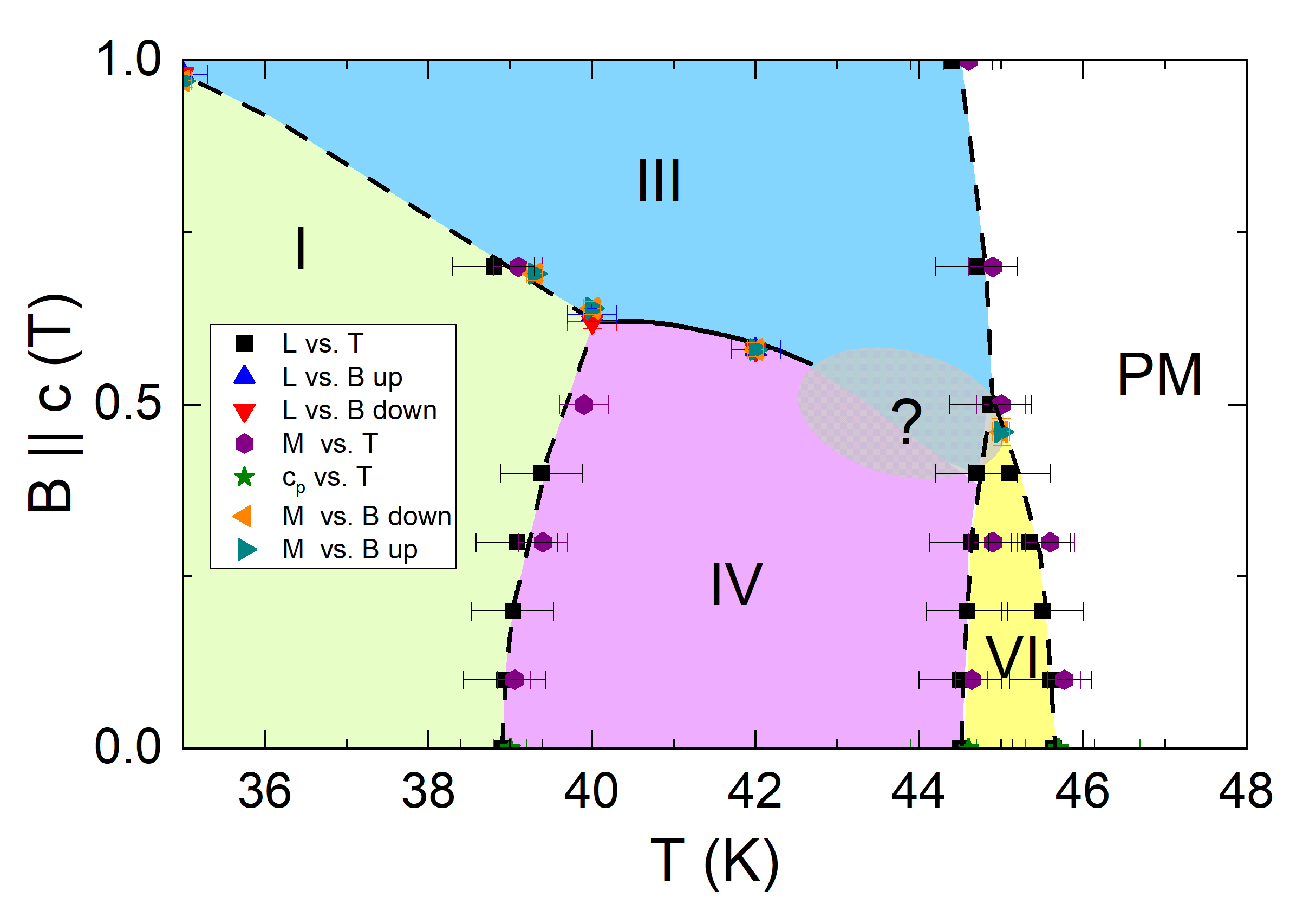}
    \caption{\rk{Magnetic phase diagram for $B||c$ in more detail around \tn\ (see Fig.~\ref{fig:phasediagram}). The exact positions of the boundaries between phases III, IV, and VI at $B||c\simeq 0.5$~T cannot be precisely determined by our data.}}
    \label{fig:phasediagram_zoom}

\end{figure}

Applying a magnetic field parallel to the $c$ axis induces significant changes to the magnetic phases as shown by the many anomalies in the thermal expansion coefficients \alc\ and Fishers' specific heat at various fields displayed in Fig.~\ref{fig:TE-field}. In addition to the measurements at constant magnetic field, isothermal magnetostriction and magnetisation data are shown in Figs.~\ref{fig:MS-MB-all} and \ref{fig:lambda_dMdB}. The observed anomalies are used to construct the magnetic phase diagram in Figs.~\ref{fig:phasediagram} and \ref{fig:phasediagram_zoom}. A detailed analysis and discussion of the anomalies will be given in §~\ref{sec:skyrmion} and \ref{sec:discussion}.

As shown in §~\ref{zero}, in zero magnetic field three distinct magnetic phases are observed: phase I at $T \leq$~\ttwo, phase IV  at \ttwo~$\leq T\leq$~\tone\ and phase VI between \tone\ and \tn\ (labelling of phases is done in consistency with Ref.~\cite{Khanh.2022}). Phase I consists of a double-\textit{Q} constant moment solution, with a helix and orthogonal spin density wave propagating on the principal magnetic propagation vectors \textbf{q}$_1$ and \textbf{q}$_2$; a second helix propagates along \textbf{q}$_1$ + 2\textbf{q}$_2$, which connects the arms of the star providing a constant moment solution \cite{Wood.2023, Paddison.2024}. Phase IV has either a sinusoidal or a helical spin structure~\cite{Paddison.2024}. The spin structure of the phase VI is unknown. For small magnetic fields $B \parallel c$, \tn\ decreases while \tone\ increases. Both phase boundaries merge at around $0.5$\,T thereby closing phase VI \luk{as summarised in Fig.~\ref{fig:phasediagram_zoom}. As indicated in the this figure, the detailed slope of the phase boundaries around $B\simeq0.5$\,T and $T\simeq 45$~K, separating phases IV, VI (and III), are not exactly clarified by our measurements since the associated anomalies are weak and partly overlap.} Similar to \tone, \ttwo\ is shifted to higher temperatures with increasing field so that phase I is stabilized over phase IV. 

At intermediate fields, anomalies in the magnetostriction and thermal expansion indicate the field-driven evolution of the square skyrmion lattice (phase II) and a fan structure (phase III). Specifically, isothermal magnetisation at $T = 42$\,K exhibits a jump at $B_{\rm IV-III} = 0.58(2)$\,T signaling a first order phase transition from phase IV to III. Similarly, dilatometric and magnetic measurements at $T \geq 20$\,K and in fields between $\sim 0.7$\,T and $\sim 2$\,T enable us to investigate the phase boundary between phases I and III. Notably, the associated anomalies qualitatively change with decreasing temperature which implies that the nature of the phase transition changes from a continuous character to a discontinuous one upon cooling. This is seen in $\lambda_{\rm c}$ and $\partial M_{\rm c}/ \partial B$ which show a $\lambda$-like behaviour at 35~K and 40~K (see Fig. \ref{fig:lambda_dMdB}). Concomitantly, $\lambda$-like anomalies are also visible in $\alpha_{c}$ and $\partial(\chi T)/\partial T$ at 1~T and 0.7~T (Fig.~\ref{fig:TE-field} \luk{and Fig.~\ref{fig:tricrit}a,b}). With decreasing temperature however the $\lambda$-like character of the magnetostriction anomaly vanishes and becomes more symmetric until it is undoubtedly symmetric around $25$\,K (Fig.~\ref{fig:lambda_dMdB}a). In the temperature-dependent measurements, the anomaly in \alc\ looses its $\lambda$-shape between $1$\,T and $1.3$\,T thereby further validating the observation in the isothermal studies. The symmetric peaks imply jumps in the $c$ axis length and in the magnetisation, \textit{i.e.}, they prove the first order nature of the phase transition in this region of the phase diagram. Our observations hence imply the presence of a tricritical point on the phase boundary between phase I and III at $\sim 33$\,K. 

\begin{figure}[htb]
    \centering
    \includegraphics[width=1\columnwidth,clip]{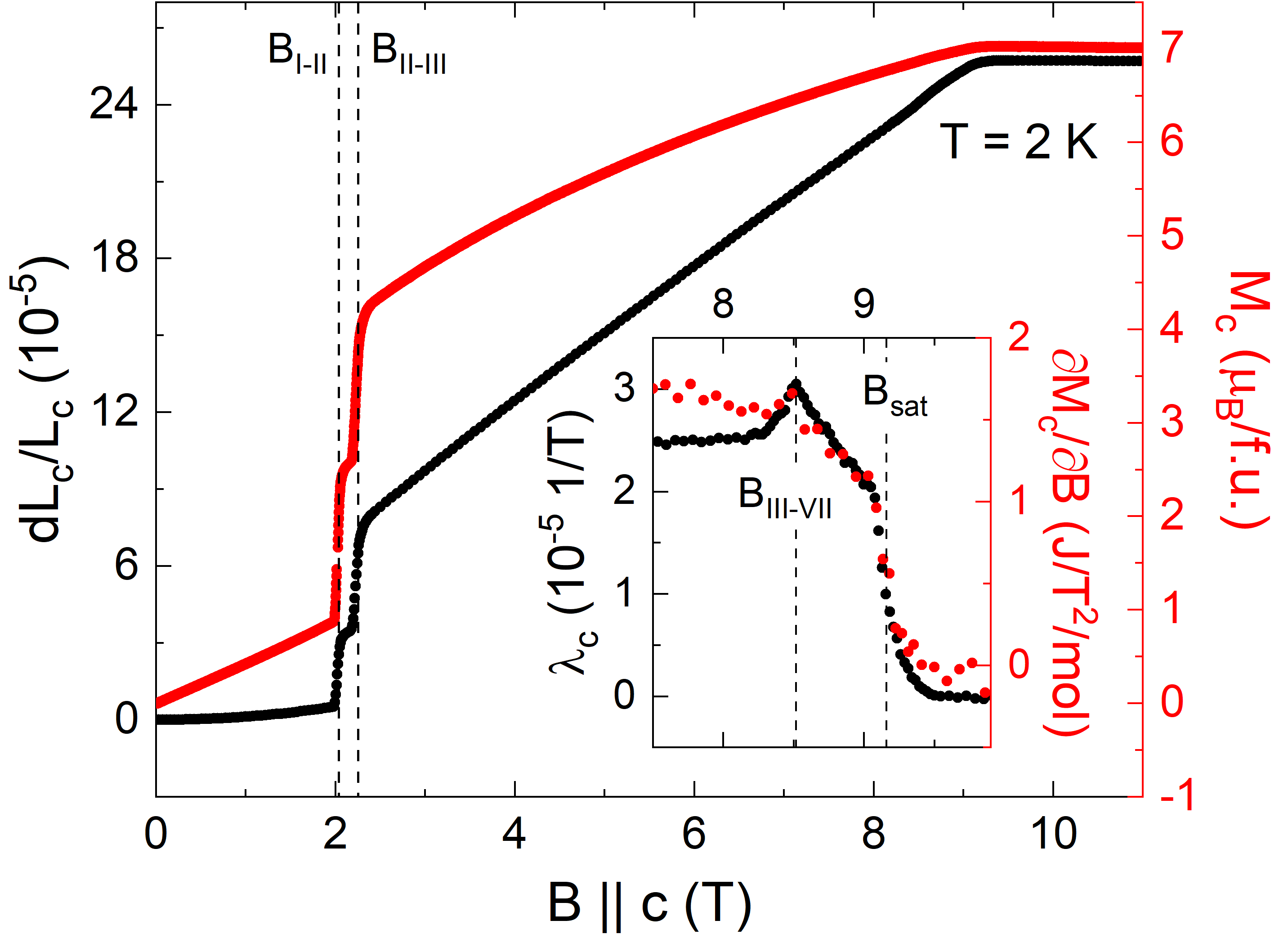}
    \caption{Magnetostriction $dL_{\rm c}(B)/L_{\rm c}$ (left ordinate) and isothermal magnetisation $M_{\rm c}$ (right ordinate) as a function of magnetic field $B \parallel c$ at $T = 2$\,K. Inset: Magnetostriction coefficient $\lambda_{\rm c}$ (left ordinate) and magnetic susceptibility $\partial M_{\rm c}/ \partial B$ (right ordinate) around the saturation field $B_{\rm sat}$. Vertical dashed lined show the transitions into and out of the skyrmion lattice phase at \bskyrin\ and \bskyrout, $B_{\rm III-VII}$, and \bsat.}
    \label{fig:MS-MB-2K}
\end{figure}

Below about 20\,K, the evolution of the skyrmion lattice phase II is evidenced by the appearance of two subsequent jumps in magnetostriction and isothermal magnetisation signaling the discontinuous phase boundaries I-II and II-III. The behaviour at $T = 2$\,K is shown in detail in Fig.~\ref{fig:MS-MB-2K} which shows that phase II extends from \bskyrin(2\,K)$=2.02(2)$\,T to \bskyrout(2\,K)$ = 2.23(2)$\,T with  \bskyrin\ and \bskyrout\ being the critical fields of the phase transition from phase I to II and phase II to III respectively~\footnote{Similar behaviour is reported in Ref.~\cite{Prokleska.2006} where magnetostriction shows two consecutive but smeared-out anomalies. The measurement temperature in Ref.~\cite{Prokleska.2006} is not specified but the reported data suggest $T<10$~K.}. Both critical fields show hysteretic behaviour which further illustrates the discontinuous character of the phase boundaries (see Fig.~S10 and Fig.~S12 in the SM) with an field-hysteresis $\sim 0.05(1)$\,T. Upon heating, the skyrmion lattice phase is suppressed by phase III and becomes narrower for higher temperatures \luk{(see Fig.~\ref{fig:phasediagram})}.

The transition temperature from the PM phase into the fan structure phase III is continuously suppressed in external magnetic fields as expected for an antiferromagnetically ordered state. In contrast to previously reported phase diagrams~\cite{Khanh.2022, Yasui.2020, Wood.2023}, phase III however does not extend to highest field since we observe a novel phase VII below about $15$\,K. The phase boundary III-VII is nearly temperature-independent as evidenced by a peak in $\lambda_{\rm c}$ and $\partial M_{\rm c}/ \partial B$ at  $B_{\rm III-VII}$~$\simeq 8.5$\,T. The observed symmetric anomaly (see inset Fig.~\ref{fig:MS-MB-2K}) indicates a first order phase transition as also illustrated by the small hysteresis as shown in Fig.~S10 and Fig.~S12 in the SM. The respective jumps at $T=2$~K in $\Delta L_{\rm c}/L_{\rm c}$ and $\Delta M$ amount to $1.9(5) \times 10^{-6}$ and $3.9(9)\times 10^{-3}$~\mbfu , respectively. Finally, at 2~K the fully polarised phase is reached above $\sim 9.3$\,T (cf. Fig~\ref{fig:phasediagram}). 

\subsection{Uniaxial pressure effects on the skyrmion lattice Phase}\label{sec:skyrmion}

The response of the skyrmion lattice phase to uniaxial pressure can be deduced from the jump-like features in the data and quantified via Clausius-Clapeyron equations (e.g.,~\cite{Stockert2012}). For constant temperature the equations yield the dependence of the transition field $B^*$ on uniaxial pressure applied along the $i$ axis from the associated jumps in the length $\Delta L_i$ and in the magnetisation $\Delta M_{B||i}$:

\begin{equation} \label{eq:CC_T_const} 
\left.\frac{{\partial}B^*}{{\partial}p_i}\right|_T = V_{\mathrm{m}}\frac{\Delta L_i/L_i}{\Delta M_{B||i}}.
\end{equation}

Analogously, when measuring at constant magnetic field $B$, the ratio of the jumps in the length $\Delta L_i$ and the entropy jump $\Delta S$ yield the uniaxial pressure dependence of the ordering temperature $T^*(B)$: 

\begin{equation}
\left.\frac{\partial T^*} {\partial p_i} \right|_B = V_{\mathrm{m}}\frac{\Delta L_i/L_i}{\Delta S} 
\end{equation}

For our analysis, we deduce the $\Delta S$ from the slope of the phase boundary at constant pressure at the respective magnetic field by exploiting $\Delta S = -  \Delta M_{B||i} / (\left. {\partial}T^*/{\partial}B\right|_p) $. This finally yields:

\begin{equation} \label{eq:CC_B_const} 
\left.\frac{{\partial}T^*}{{\partial}p_i}\right|_B = - V_{\mathrm{m}} \left.\frac{{\partial}T^*}{{\partial}B}\right|_p \frac{\Delta L_i/L_i}{\Delta M_{B||i}}
\end{equation}

Similarly to the above mentioned procedure to determine anomaly sizes, we have extracted the jumps $\Delta L_c$ and $\Delta M_{B||c}$ from the experimental data by fitting lines to the data well below and above the anomalies. The resulting jumps and calculated uniaxial pressure dependencies obtained from the isothermal measurements at the phase boundaries into the SKL phase (I-II, \textit{i.e.}, $B_{\rm I-II}$) and out of the SKL phase (II-III; $B_{\rm II-III}$) are displayed in Table~S1. The calculated uniaxial pressure dependencies for the respective phase boundaries exhibit no significant changes with temperature, being $\partial B_{\rm I-II} / \partial p_{\rm c} \simeq 0.17$\,T/GPa and $\partial B_{\rm II-III} / \partial p_{\rm c} \simeq 0.24$\,T/GPa on average.

Furthermore by extracting the slope of the respective phase boundary $\partial T_{\rm j} / \partial B$ (j = I-II, II-III) the pressure dependency of the transition temperature $\partial T_{\rm j} / \partial p_{\rm c}$ can be calculated using Eq.~\ref{eq:CC_B_const} and values from Table~S1 in the SM. Specifically, $\partial T_{\rm j} / \partial B$ was approximated by fitting a polynomial to the respective phase boundary and determining the field derivative. Table~S2 in the SM lists the approximated slopes of the respective phase boundaries and the obtained uniaxial pressure dependencies. For both phase boundaries I-II and II-III the uniaxial pressure dependencies of the transition temperatures $T_{\rm I-II}$ and $T_{\rm II-III}$ are positive being on the order of 10\,K/GPa. Furthermore, for both boundaries, $\partial T_{\rm j} / \partial p_{\rm c}$ increases for higher magnetic fields which can be mostly attributed to the increasing slope of the phase boundary $\partial T_{\rm j} / \partial B$.


\subsection{Discussion}\label{sec:discussion}

The phase diagram in Fig.~\ref{fig:phasediagram} shows the presence of six distinct ordered phases evolving below \tn\  in external magnetic fields $B$ applied along the crystallographic $c$ axis. Similar to the analysis of thermal expansion and magnetostriction anomalies at the boundaries of the SKL phase, the data in Figs.~\ref{fig:TE-0T}-\ref{fig:lambda_dMdB} as well as further data presented in the SM~\cite{supplement} imply the dependencies of the respective ordering phenomena upon application of uniaxial pressure along the $c$ axis ($p_{\rm c}$). The results are summarized in table~\ref{tab:pressuredep} and, interestingly, show only positive values for the uniaxial pressure dependencies. This implies that phase I is stabilised over all adjacent phases II, III and IV in temperature and magnetic field. Phases IV and VI, the remaining zero field phases, are both shifted to higher temperatures \luk{(see Fig.~\ref{fig:phasediagram_zoom}.)} and span over a larger temperature interval for $p||c$ since $\partial T_{\rm I-IV} / \partial p_{\rm c} < \partial T_{\rm IV-VI} / \partial p_{\rm c} < \partial T_{\rm VI-PM} / \partial p_{\rm c}$. Furthermore, phase IV is stabilised over phase III under uniaxial pressure $p_{\rm c}$. As already discussed above, the skyrmion lattice phase shifts to higher fields and widens in magnetic field. Also phases III and VII are stabilised towards higher temperatures and magnetic fields upon applying $p||c$ axis. Notably, phase VII is particularly sensitive to $p_{\rm c}$ as $\partial B_{\rm III-VII}/\partial p_{\rm c}$ is two orders of magnitude larger than the pressure dependencies of all other phase boundaries. It is of the order of $5$\,T/GPa, implying that phase VII is strongly suppressed in favour of phase III.

Pressure effects on the skyrmion lattice phase are of particular interest as they provide further insight into the microscopic mechanism stabilising this phase. The observed uniaxial pressure dependencies of the phase boundaries enclosing the skyrmion lattice phase are all positive, \textit{i.e.}, uniaxial pressure along the $c$ axis stabilizes the skyrmion lattice towards higher fields and temperatures. Furthermore, for all measured temperatures $\partial B_{\rm II-III} / \partial p_{\rm c}$ is larger than $ \partial B_{\rm I-II} / \partial p_{\rm c}$ so that the skyrmion lattice phase also widens in magnetic field at an approximate rate of $\Delta B_{\rm skyr}/p_{\rm c} \approx 0.07$\,T/GPa. An enhancement of the skyrmion lattice phase under uniaxial pressure is also observed in materials such as Gd$_2$PdSi$_3$ \cite{Spachmann.2021}, Cu$_2$OSeO$_3$ \cite{Levatic.2016} and MnSi \cite{Chacon.2015}. Especially interesting for comparison is Gd$_2$PdSi$_3$ which also crystallises in a centrosymmetric structure and is predominantly governed by RKKY interactions~\cite{Kurumaji.2019}. Similar to the present case the SKL phase widens in magnetic field in Gd$_2$PdSi$_3$ for uniaxial pressure parallel to the $c$ axis at approximately half the rate compared to \GRS. However for Gd$_2$PdSi$_3$ the sign of pressure dependence is opposite so that the SKL phase shifts towards lower magnetic fields. The pressure dependence of the critical fields is weaker by approximately one order of magnitude in Gd$_2$PdSi$_3$ than in \GRS~\cite{Spachmann.2021}. In contrast to the strongly differing pressure dependencies of the critical fields, the ordering temperatures of the SKL phases in both materials change is of similar magnitude of however again opposite signs: While in Gd$_2$PdSi$_3$, $\partial T_{\rm skyr} / \partial p_{\rm c} = -6.1$\,K/GPa, in \GRS\ we find $\partial T_{\rm skyr} / \partial p_{\rm c} \simeq +15$\,K/GPa (see table~\ref{tab:pressuredep}). In case of the two non-centrosymmetric systems Cu$_2$OSeO$_3$ and MnSi the respective SKL phases also widen in magnetic field under pressure \cite{Chacon.2015, Levatic.2016}. Moreover, the SKL phase in MnSi shows the general shift towards higher fields, while for Cu$_2$OSeO$_3$ the SKL phase extends towards higher temperatures as observed for \GRS. 

\begin{table}[b]
\centering
\caption{Uniaxial pressure dependencies of the phase boundaries $j$ shown in Fig. \ref{fig:phasediagram} for $p||c$. The presented values are from calculations using the Clausius-Clapeyron and Ehrenfest relations (see Eq.~\ref{eq:Ehrenfest_2}, Eq.~\ref{eq:CC_B_const} and Eq.~\ref{eq:CC_T_const}). Values marked with an asterisk have been obtained from phase boundaries where more than one value could be calculated. In these cases, the averages are taken and the error bars include the variations at the phase boundary. If no quantitative values can be obtained the sign of the pressure dependence is given.}
\begin{tabular}{c| c| c}
	  \hline \hline
   phase boundary $j$~ & $ ~ \partial B_{\rm j}/ \partial p_{\rm c}$  [T/GPa] ~& $ ~\partial T_{\rm j} / \partial p_{\rm c}$  [K/GPa]~  \\
\hline
  I - IV    &  +   &   1.2(3)  \\
  IV - VI   &  +    &   2.6(4)  \\
  VI - PM   &  +    &   3.3(3)  \\
  I - III (1$^{st}$-order)   &  0.20(4)*   &   4.5(6) \\
  I - III (2$^{nd}$-order)   &  0.13(3)*   &   + \\
  IV - III   &  0.11(2)   &   + \\
  III - PM  &    +       &   +  \\
  I - II    &    0.17(3)*       &   20(-13,+60)*  \\
  II - III  &  0.24(3)*   &   15(-11,+40)*  \\
  III - VII &    4.3(1.5)       &  +   \\
  VII - PM  &    +       &     +    \\

  \hline \hline
 
\end{tabular}
\label{tab:pressuredep}
\end{table}

While uniaxial pressure effects on all ordered phases are always positive, the effect of external magnetic fields $B||c$ differs for the various phases not only quantitatively but also with respect to its sign. At low magnetic field $B\lesssim 0.5$~T phase I and IV are stabilised over their respective higher temperature phases, \textit{i.e.}, the phase boundaries show positive slope $\partial T_1/\partial B_{\rm c} >0$ and $\partial T_2/\partial B_{\rm c} >0$ \luk{(see Fig.~\ref{fig:phasediagram_zoom})}. Thermodynamically, this is associated with an increase of magnetisation or its derivative upon cooling at the respective phase boundaries which is indeed observed in the magnetisation data in Fig.~\ref{fig:chi-T}. All other phase boundaries show negative slopes, \textit{i.e.}, the underlying magnetic orders are suppressed by $B||c$ (see Fig.~\ref{fig:phasediagram}).

Our results show that the previously unreported phase VII forms a pocket at high-fields and low-temperatures with an upper boundary to the fully polarized state, $B_{\rm sat}(T)$ is very similar to what would be expected for phase III (\textit{cf.} Fig.~\ref{fig:phasediagram}). While the phase boundary $B_{\rm III-VII}(T)$ is nearly temperature independent, it displays a comparably giant uniaxial pressure dependence so that phase VII will be fully suppressed by applying $p_{\rm c}$ of a few tenth of GPa. By means of the Clausius-Clapeyron equation we obtain an upper limit of the entropy changes of $\Delta S_{\rm III-VII}(T=2~{\rm K})<5\times 10^{-4}$~\jmk\, which uses the observed jump in $M$ and the tiny slope ($\ll0.1$~T/K) of the phase boundary at the respective transition. Although the spin structure of the new phase has not been studied yet, first principal numerical studies by Bouaziz \etal~\cite{Bouaziz.2022} may reveal a possible nature of this phase. Their calculations for \GRS\ predict that in the fully polarized state close to the boundary of the cycloidal phase (phase III) single metastable skyrmions emerge. One might speculate whether such skyrmions form a superstructure which could be our observed phase VII. If so, it is very sensitive to and easily suppressed by pressure $p_{\rm c}$ while entropically it is very similar to phase III.


\begin{figure}[h!]

    \centering
    \includegraphics[width=0.9\columnwidth,clip]{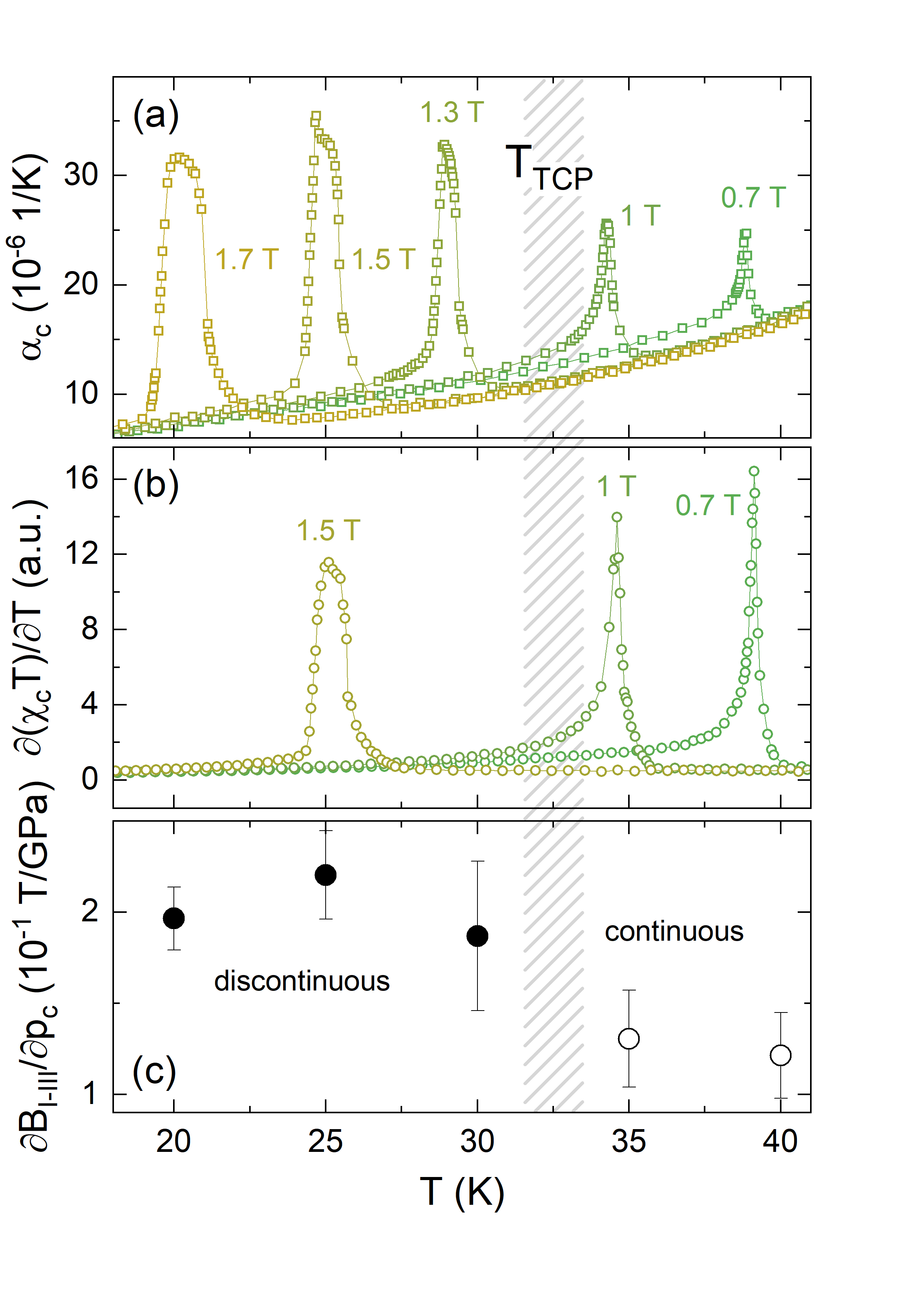}
    \caption{\luk{Anomalies associated with the phase transition between phases I and III in (a) the linear thermal expansion coefficient $\alpha_{\rm c}$, and (b) Fisher's specific heat $\partial (\chi_{\rm c} T)/\partial T$ }. (c) Uniaxial pressure dependence $\partial B_{\rm I-III} / \partial p_{\rm c}$ at the phase boundary I-III. The dashed region marks the temperature regime of the tricritical point.}
    \label{fig:tricrit}

\end{figure}

We finally discuss the phase boundary between phase I and III which, as mentioned above, displays a continuous nature above $35$\,K as demonstrated by textbook-like anomalies at $B_{\rm I-III}$ while we observe 1$^{st}$ order discontinuities in $M$ and $L$ below 25~K (see Fig.~\ref{fig:TE-field} and \ref{fig:MS-MB-all}). \luk{The evolution of the anomaly shape along the phase boundary is illustrated Fig.~\ref{fig:tricrit}a and \ref{fig:tricrit}b which shows the change from asymmetric and clearly $\lambda$-shaped anomalies (confirming the continuous nature of the transition) for $T>T_{\rm TCP}$ to rather symmetric anomalies with sharp low-temperature flanks (indicating the 1$^{st}$ order character of the boundary) in $\alpha_{\rm c}$ and $\partial (\chi_{\rm c}T)/\partial T$ for $T<T_{\rm TCP}$ to .} Our findings suggest the existence of a tricritical point (TCP) at $T_{\rm TCP}\simeq 33$\,K (see Fig.~\ref{fig:phasediagram}), \textit{i.e.}, well above the triple point where the SKL evolves in between phases I and III. The existence of a TCP is further supported by the temperature dependence of the uniaxial pressure dependence of $B_{\rm I-III}(T)$ presented in Fig.~\ref{fig:tricrit}c. Above and below $T_{\rm TCP}$ the uniaxial pressure dependence of the critical field $\partial B_{\rm I-III} / \partial p_{\rm c}$ is fairly constant but its value changes by about 40\% at $T_{\rm TCP}$. Overall, the continuous transition between double-\textit{Q} magnetic structures in phases I and III evolves a discontinuous character upon cooling \luk{(see Fig.~\ref{fig:tricrit}a,b)} before the SKL phase evolves upon further cooling at $T_{\rm tri}\simeq 18\,\rm{K}<T_{\rm TCP}$. This behaviour shows slight similarities but essentially contrasts to findings in the skyrmion lattice systems MnSi~\cite{Bauer.2013, Zhang.2015}, Cu$_2$OSeO$_3$ \cite{Chauhan.2019} and GaV$_4$S$_8$~\cite{Liu.2020} where tricritical behaviour is reported only at triple points of the phase diagram. In all three examples, the TCP appears at a phase boundary towards the fully polarised phase which in case of the skyrmion-pocket phases in MnSi and Cu$_2$OSeO$_3$ does not involve a phase boundary towards the skyrmion phase while only in GaV$_4$S$_8$~\cite{Liu.2020} the TCP edges the SKL phase.

\section{Summary}

We report high-resolution capacitance dilatometry, specific heat, and magnetisation studies which are used to complete the magnetic phase diagram in \GRS . We observe three successive antiferromagnetic phases in zero magnetic field (phases I, IV, VI), with phase VI not yet reported and of unknown structure. In addition, we also find a new high-field phase (phase VII), which features a comparably giant uniaxial pressure dependence. By means of our dilatometric data we determine magneto-elastic effects as well as uniaxial pressure dependencies of the various phases. The skyrmion lattice phase is enhanced towards higher fields and temperatures and widens at a rate of 0.07~T/GPa when uniaxial pressure is applied along the $c$ axis. Notably, the SKL pocket phase evolves through a triple point $T_{\rm tri}$ from a phase boundary between the double-\textit{Q} magnetic structures in phases I and III which in addition \luk{indicates a} tricritical point at  $T_{\rm TCP}\simeq 33\,\rm{K}>T_{\rm tri}\simeq 18$\,K, thereby highlighting the relevance of critical fluctuations for the evolution of the skyrmion lattice phase in \GRS . \\

\begin{acknowledgements}
We acknowledge valuable experimental advise for the specific heat studies at the University of Warwick by Prof. Martin Lees. Support by Deutsche Forschungsgemeinschaft (DFG) under Germany's Excellence Strategy EXC2181/1-390900948 (the Heidelberg STRUCTURES Excellence Cluster) is gratefully acknowledged. L.~G. acknowledges funding by the International Max-Planck Research School for Quantum Dynamics (IMPRS-QD) Heidelberg. The work at the University of Warwick was supported by EPSRC, UK through Grants EP/T005963/1 and EP/N032128/1. 
\end{acknowledgements}

\bibliography{literature3}

\appendix

\renewcommand{\thefigure}{S\arabic{figure}}
\renewcommand{\thetable}{S\arabic{table}}
\renewcommand{\theequation}{S\arabic{equation}}

\setcounter{figure}{0}
\setcounter{table}{0}
\setcounter{equation}{0}

\clearpage

\section*{Supplemental information}

The supplemental information contains:

\begin{itemize}
    \item Tables: Extracted values of the jumps in isothermal magnetisation $M_{\rm c}$, magnetostriction $dL_{\rm c}/L_{\rm c}$, the respective slopes of the phase boundary $\partial T/\partial B$ and the calculated pressure dependencies $\partial B_{\rm crit}/\partial p_{\rm c}$ and $\partial T_{\rm crit}/\partial p_{\rm c}$.

    \item Scaling of the specific heat data with high temperature resolution. 
    
    \item Inverse magnetic volume susceptibility including a Curie-Weiss-fit.

    \item Specific heat and fit of the phononic contribution by means of an Einstein-Debye model.  

    \item Determination of the jumps in the specific heat and the thermal expansion coefficient.

    \item Thermal expansion and magnetostriction measurements in detail for $B||c$ as well as their respective derivatives, \textit{i.e.}, the thermal expansion and magnetostriction coefficients.

    \item Statistic magnetic susceptibility $\chi_{\rm c}$ and isothermal magnetisation $M_{\rm c}$ measurements in detail for $B||c$ as well as Fisher's specific heat $\partial (\chi_{\rm c}T)/\partial T$ and the magnetic susceptibility $\partial M_{\rm c}/\partial B$.

    \item  Determination of the jumps in magnetostriction at the phase transitions in and out of the skyrmion lattice phase.

\end{itemize}

\begin{table*}[htb]
\centering
\caption{Jumps in the isothermal magnetisation $M_{\rm c}(B)$ and the magnetostriction $dL_{\rm c}(B)/L_{\rm c}$ (see text as well as Fig.~\ref{SM:jump_MS_2K} in the SM) at $B_{\rm j}$ (j =  I-II, II-III) for various temperatures $T$  as well as the resulting uniaxial pressure dependencies calculated using the Clausius-Clapeyron equation at constant temperature (Eq.~3).} 
\begin{tabular}{c|c c c|c c c}
	  \hline \hline
   
		 T [K] & $\frac{\Delta L_{\rm c}}{L_{\rm c}}$ [$10^{-5}$] & $\Delta M_{\rm c}$ [J/T/mol] & $\partial B_{\rm I-II} / \partial p_{\rm c}$  [T/GPa] & $\frac{\Delta L_{\rm c}}{L_{\rm c}}$ [$10^{-5}$] & $\Delta M_{\rm c}$ [J/T/mol]  & $\partial B_{\rm II-III} / \partial p_{\rm c}$  [T/GPa]  \\
		\hline
           &    &  I - II  &        &   & II - III & \\
         \hline  
	    2  &  3.0(2)   &  8.8(2)       &  0.171(12)	  &  4.2(2)   &  8.7(2)  &  0.242(13)   \\
        4  &  2.9(2)   &  8.8(3)       &  0.166(13)	  &  4.2(2)   &  8.8(3)  &  0.240(14)  \\
		6  &  3,0(2)   &  8.6(3)       &  0.175(13)	  &  4.1(2)   &  8.6(3)  &  0.239(14)  \\
        8  &  2.8(2)   &  8.4(3)       &  0.167(13)	  &  4.0(2)   &  8.4(3)  &  0.240(14) \\
        10 &  2.7(2)   &  8.2(3)       &  0.167(14)	  &  3.9(2)   &  8.2(3)  &  0.240(15)  \\
        12 &  2.6(2)   &  8.1(3)       &  0.166(14)	  &  3.8(2)   &  8.1(3)  &  0.236(15) \\
        15 &  2.5(3)   &  7.7(3)       &  0.184(24)	  &  3.4(3)   &  7.7(3)  &  0.223(22)  \\
 \hline \hline
\end{tabular}
\label{tab:skyrmion_T_const}
\end{table*}

\begin{table*}[htb]
\centering
\caption{Extracted slope from the two phase boundaries I-II and II-III (see text) at $\partial T_{\rm j}/\partial B$ (j = I-II, II-III) for various magnetic fields $B$ as well as the resulting uniaxial pressure dependencies calculated using the Clausius-Clapeyron equation at constant field (Eq.~5) and values from Table \ref{tab:skyrmion_T_const}.} 
\begin{tabular}{c c c |c c c }
	  \hline \hline
   
		 B [T] & $\partial T_{\rm I-II} / \partial B$ [K/T]  & $\partial T_{\rm I-II} / \partial p_{\rm c}$  [K/GPa] & B [T] & $\partial T_{\rm II-III} / \partial B$ [K/T]  & $\partial T_{\rm II-III} / \partial p_{\rm c}$  [K/GPa]  \\
		\hline
        &  I - II  &        &   & II - III & \\
         \hline  
	    1.85(1)  &  -45(5)     &  8.3(1.4)     &  1.93(1)	  &  -26(4)   &  5.9(1.1)     \\
        1.91(1)  &  -56(4)     &  9.2(1.3)     &  2.04(1)	  &  -32(4)   &  7.5(1.1)    \\
		1.95(1)  &  -70(5)     &  12(2)    &  2.10(1)	  &  -35(4)   &  8.5(1.2)    \\
        1.97(1)  &  -81(8)     &  14(2)       &  2.15(1)	  &  -43(4)   &  10.4(1.3)   \\
        1.99(1)  &  -102(20)   &  18(4)       &  2.18(1)	  &  -53(4)   &  12.6(1.4)    \\
        2.01(1)  &  -162(100)   &  27(17)       &  2.21(1)	  &  -85(10)  &  20(3)   \\
        2.02(1)  &  -316(150)  &  54(26)      &  2.22(1)	  &  -154(75) &  37(18)    \\
 \hline \hline
\end{tabular}
\label{tab:skyrmion_B_const}
\end{table*}

\begin{figure}[h]
    \centering
    \includegraphics[width=0.7\columnwidth,clip]{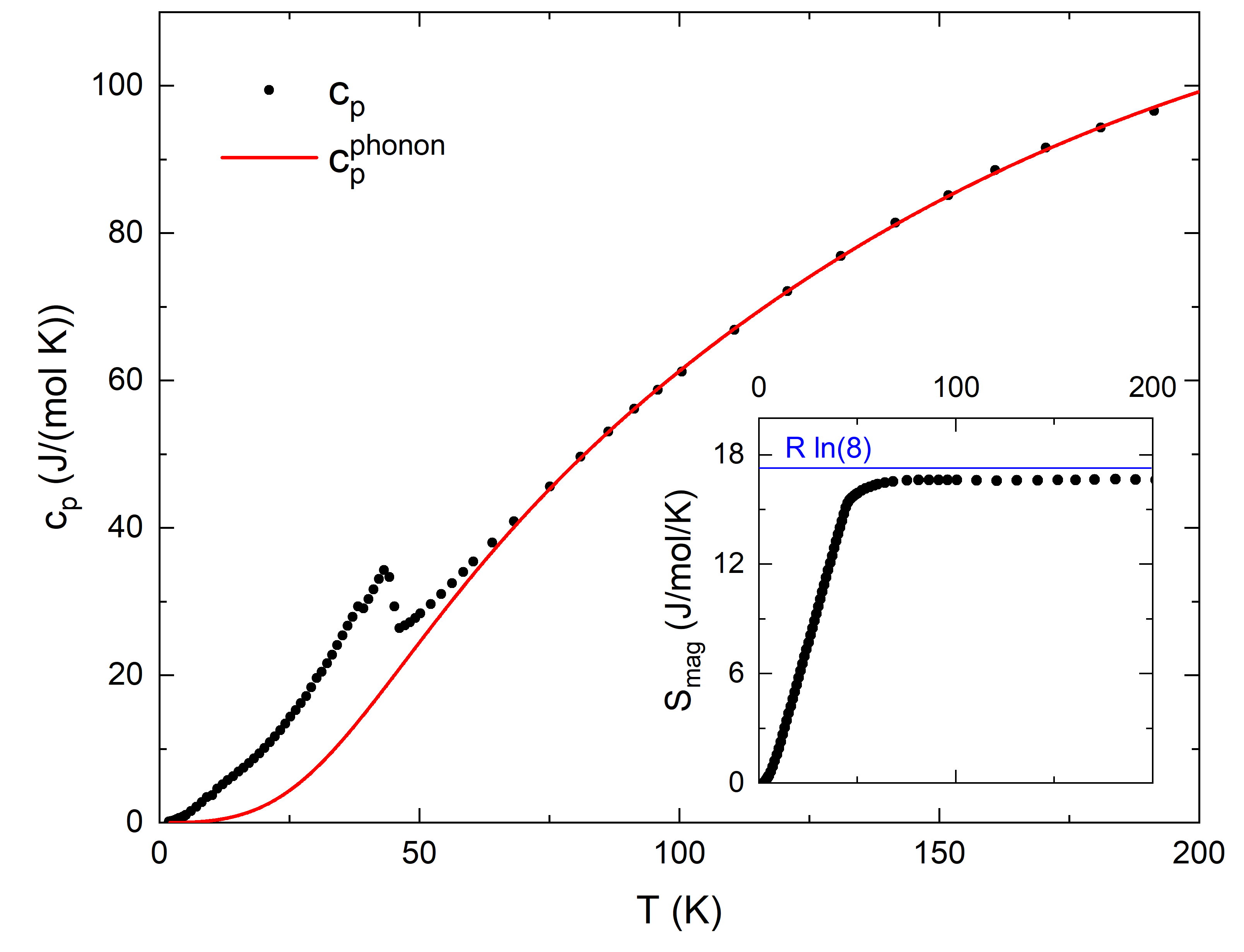}
    \caption{\luk{Specific heat $c_{\rm p}$ vs. temperature in zero magnetic field measured on a large crystal up to 300\,K. The red line represents the fitted lattice contribution $c_{\rm p}^{\rm ph}$ by means of an Einstein-Debye model. The inset depicts the magnetic entropy changes obtained by integrating $(c_{\rm p} - c_{\rm p}^{\rm ph}) / T$ which agree with the theoretically expected magnetic entropy for a Gd$^{3+}$ system (blue line).}} 
    \label{SM:cp_DE_fit}
\end{figure}

\begin{figure}[h]
    \centering
    \includegraphics[width=0.7\columnwidth,clip]{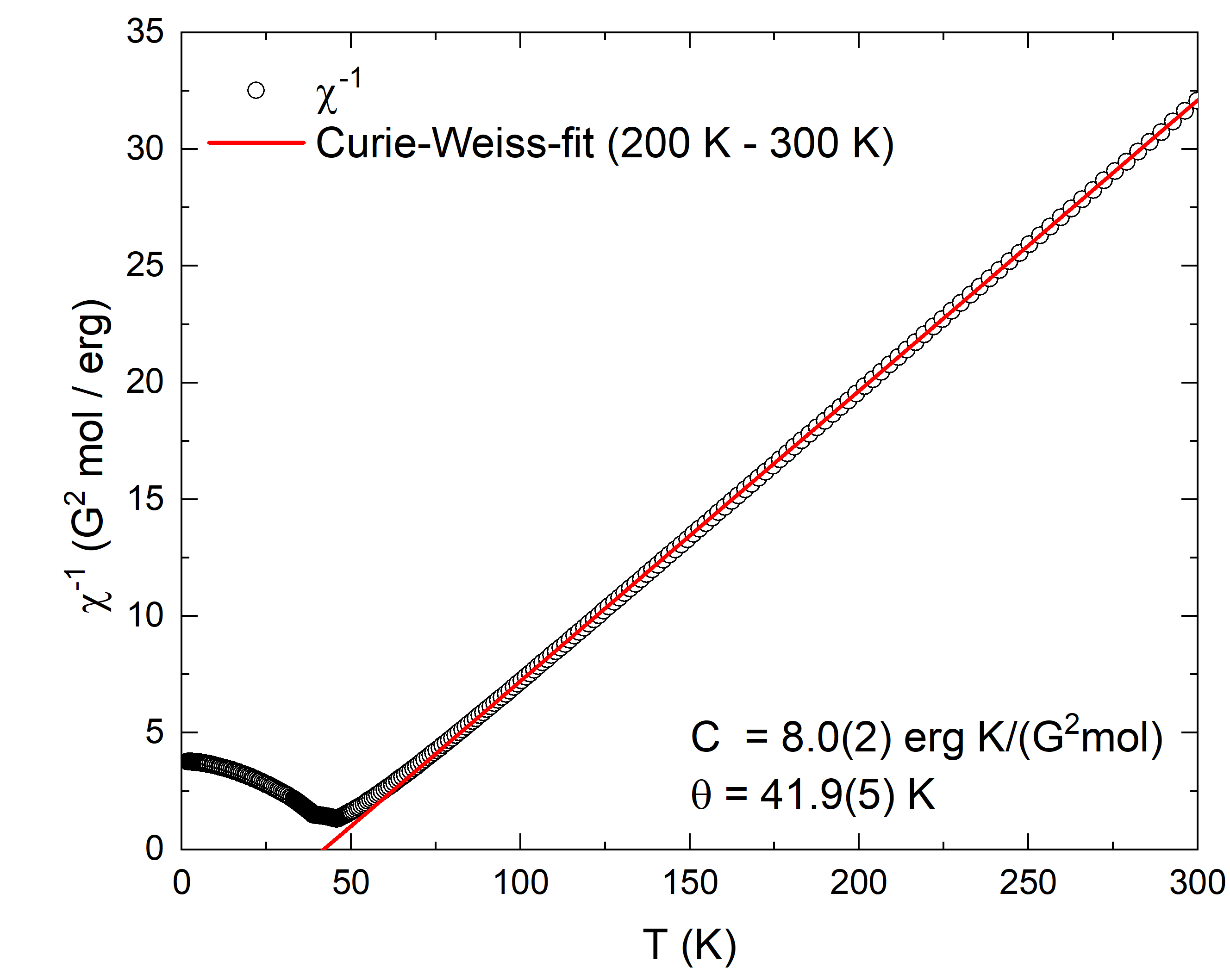}
    \caption{Inverse magnetic volume susceptibility of \GRS\ and fit by means of a Curie-Weiss law. The obtained Curie constant $C$ agrees well to the theoretically expected value of free Gd$^{3+}$. }
    \label{SM:CW}
\end{figure}

\begin{figure}[h]
    \centering
    \includegraphics[width=0.65\columnwidth,clip]{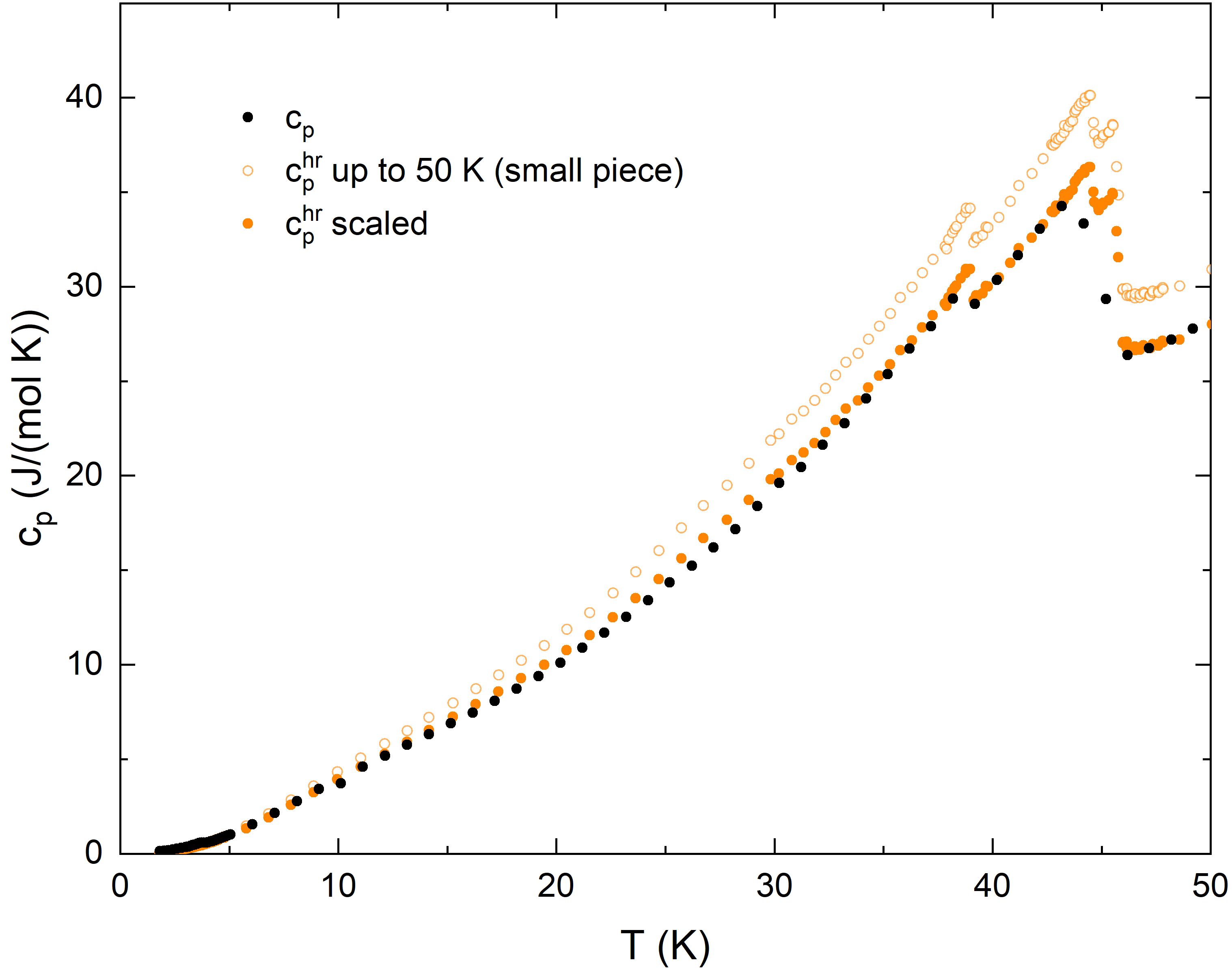}
    \caption{\luk{Specific heat $c_{\rm p}$ vs. temperature in zero magnetic field measured on a large crystal up to 300\,K (black filled dots) and on the small crystal 1 up to 50\,K with higher resolution at temperatures in the vicinities of the phase transition (open orange circles). Orange filled circles show the data of crystal 1 scaled to the large crystal data. Scaling is performed so that both measurements match above \tn .}}
    \label{SM:cp_scaled}
\end{figure}

\begin{figure}[h]
    \centering
    \includegraphics[width=0.7\columnwidth,clip]{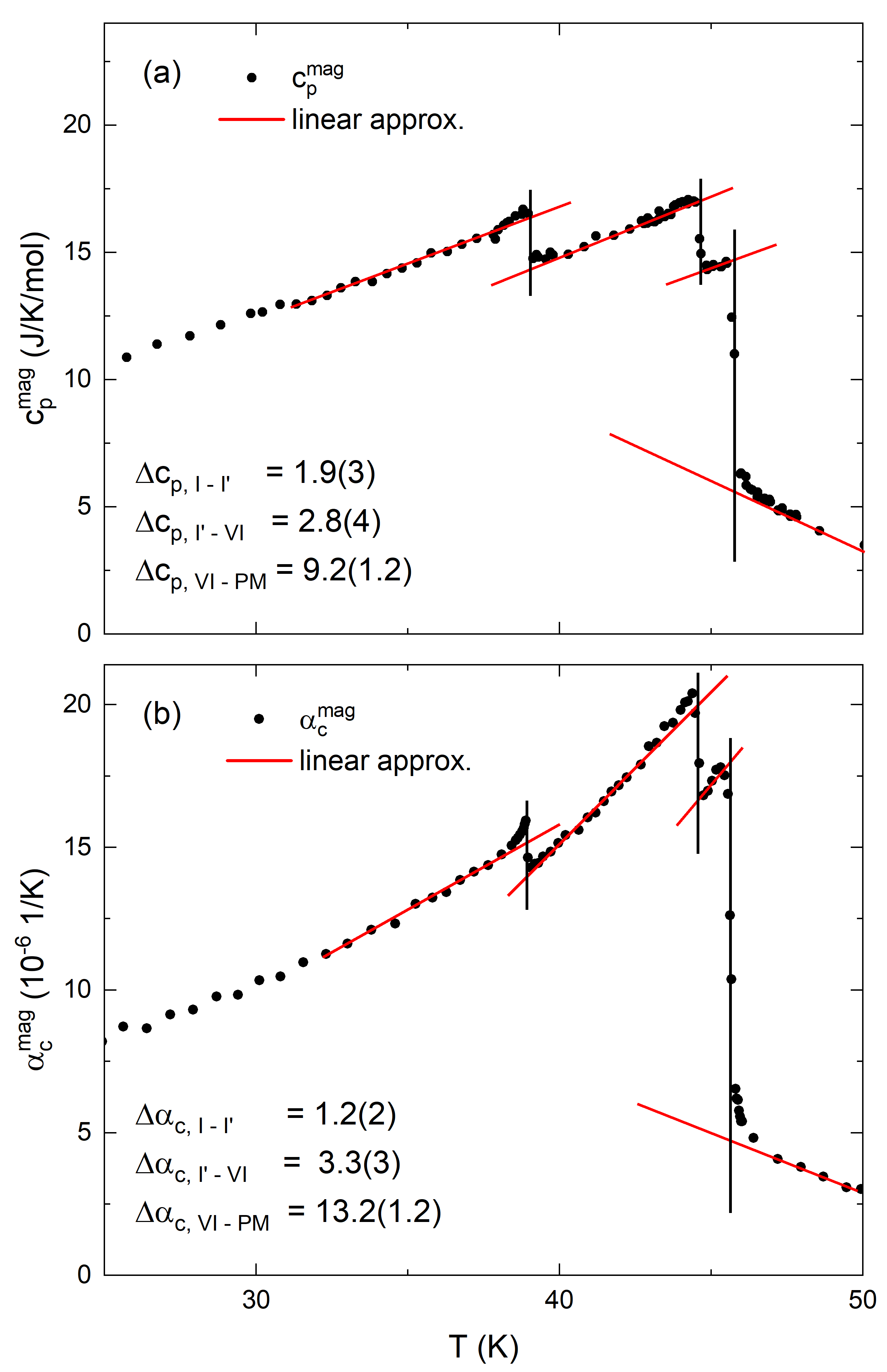}
    \caption{Determination of the jumps at the zero-field phase transitions of \GRS\ in the (a) magnetic specific heat $c_{\rm p}^{\rm mag}$ and (b) magnetic contribution to the thermal expansion coefficient $\alpha_{\rm c}^{\rm mag}$. The red lines are fits to regimes below and above the three phase transition to correct the jumps for the superimposed fluctuations at the respective phase transition.}
    \label{SM:jumps_TE_CP_0T}
\end{figure}

\begin{figure}[h]
    \centering
    \includegraphics[width=1\columnwidth,clip]{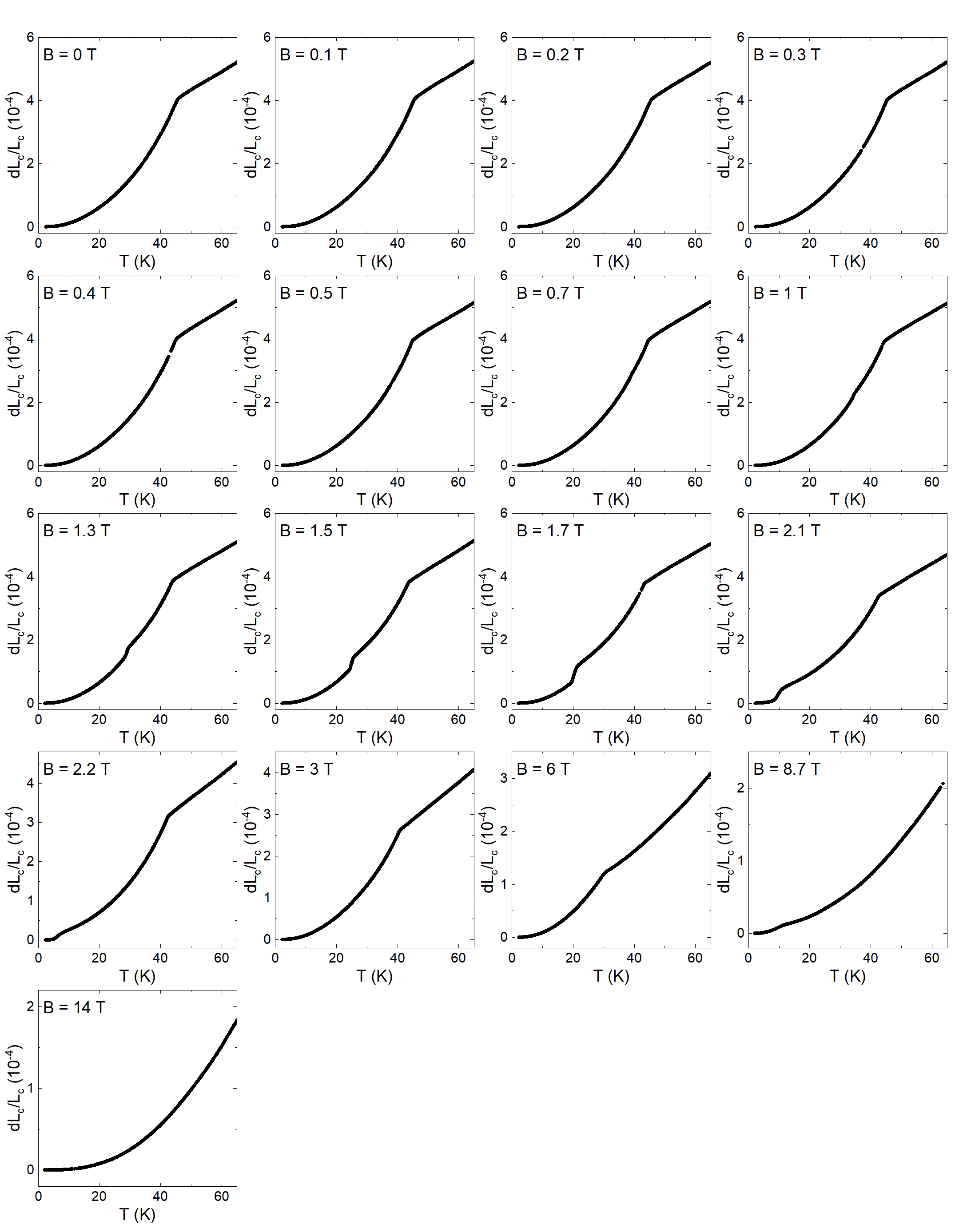}
    \caption{Thermal expansion $dL_{\rm c}/L_{\rm c}$ at various fields up to $B \parallel c = 14$\,T as a function of temperature $T$. }
    \label{SM:dLL}
\end{figure}

\begin{figure}[h]
    \centering
    \includegraphics[width=1\columnwidth,clip]{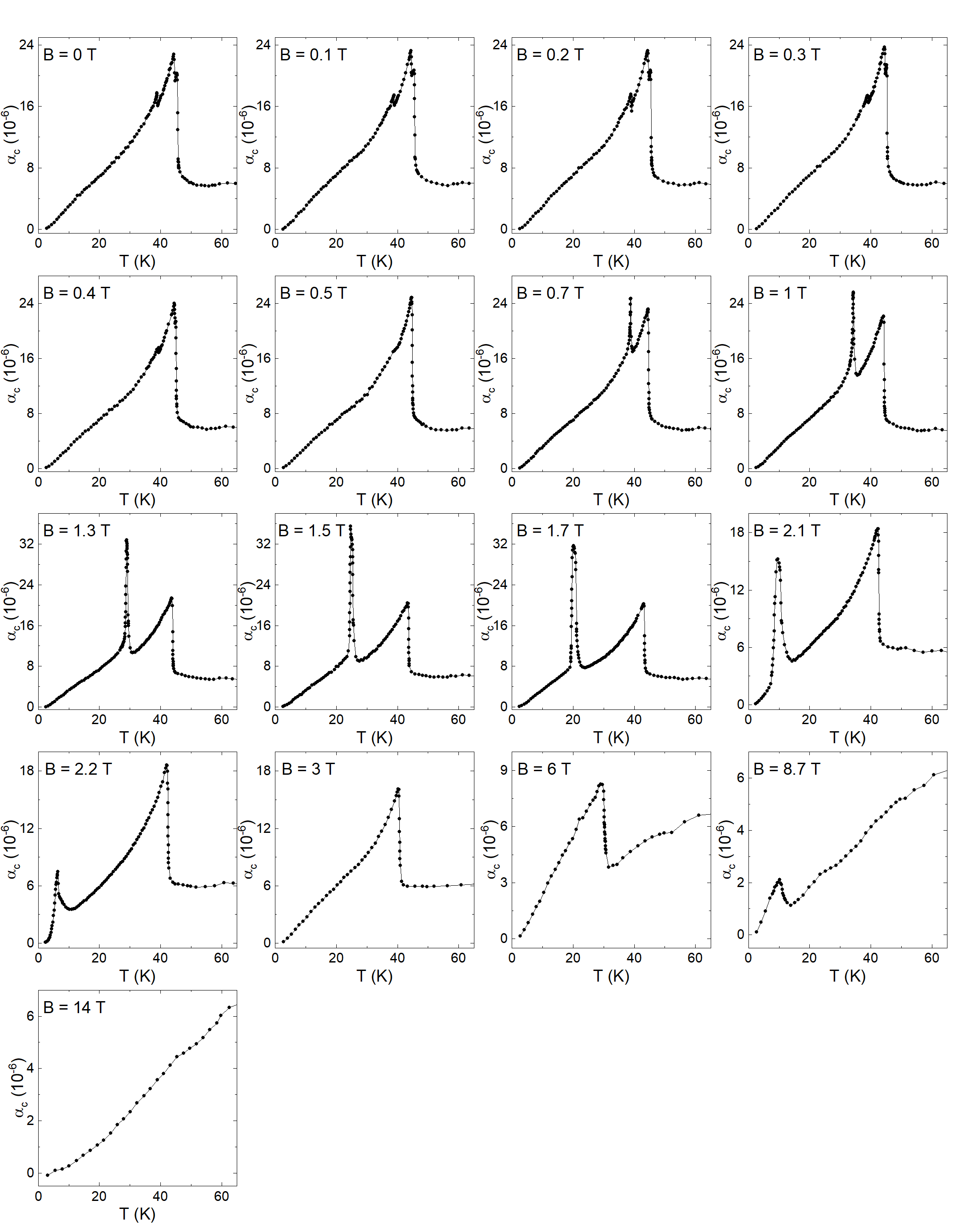}
    \caption{Thermal expansion coefficient $\alpha_{\rm c}$ at various fields up to $B \parallel c = 14$\,T as a function of temperature $T$. }
    \label{SM:alpha}
\end{figure}

\begin{figure}[h]
    \centering
    \includegraphics[width=1\columnwidth,clip]{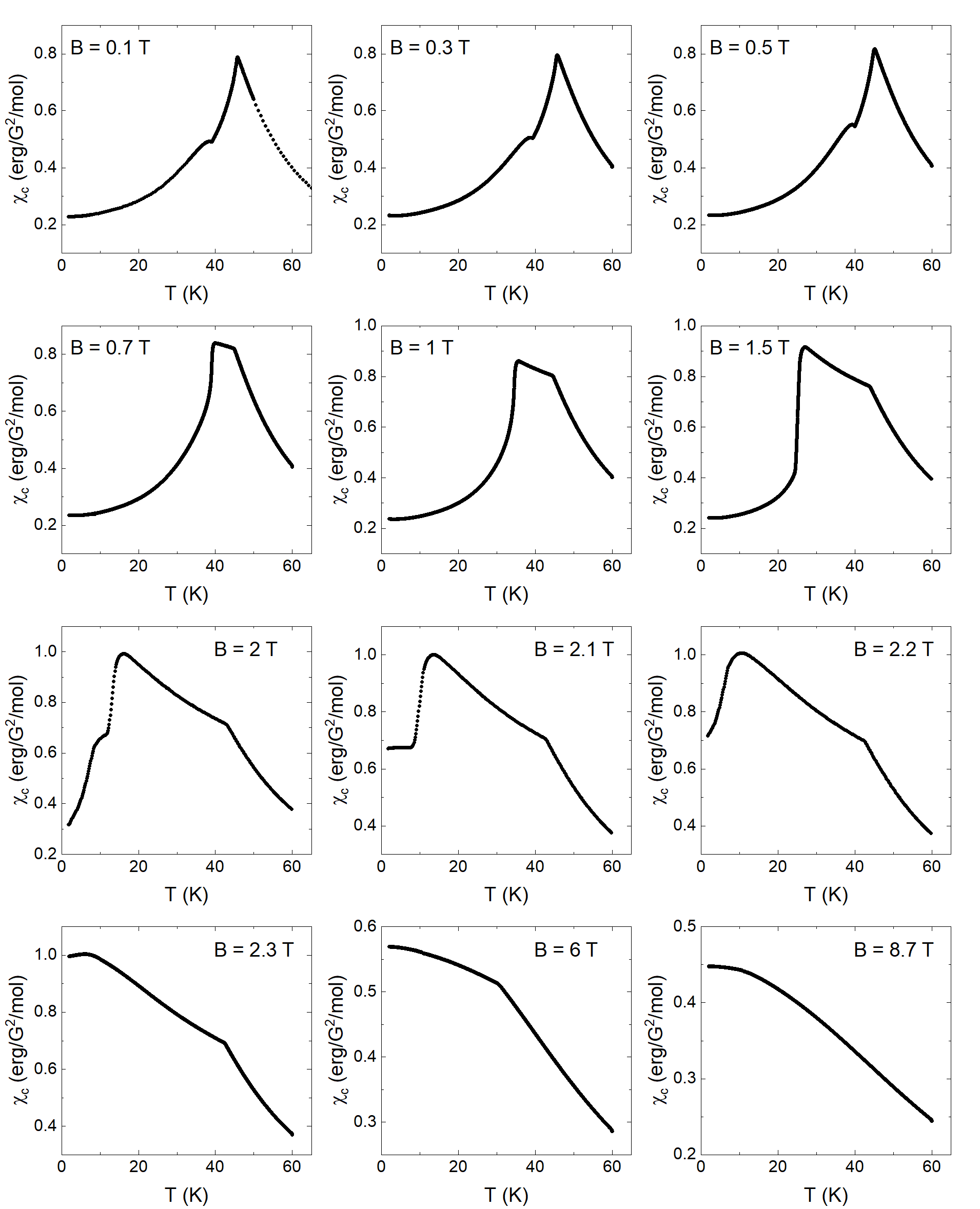}
    \caption{Static magnetic susceptibility $\chi_{\rm c}$ at various fields up to $B \parallel c = 8.7$\,T as a function of temperature $T$. }
    \label{SM:MT}
\end{figure}

\begin{figure}[h]
    \centering
    \includegraphics[width=1\columnwidth,clip]{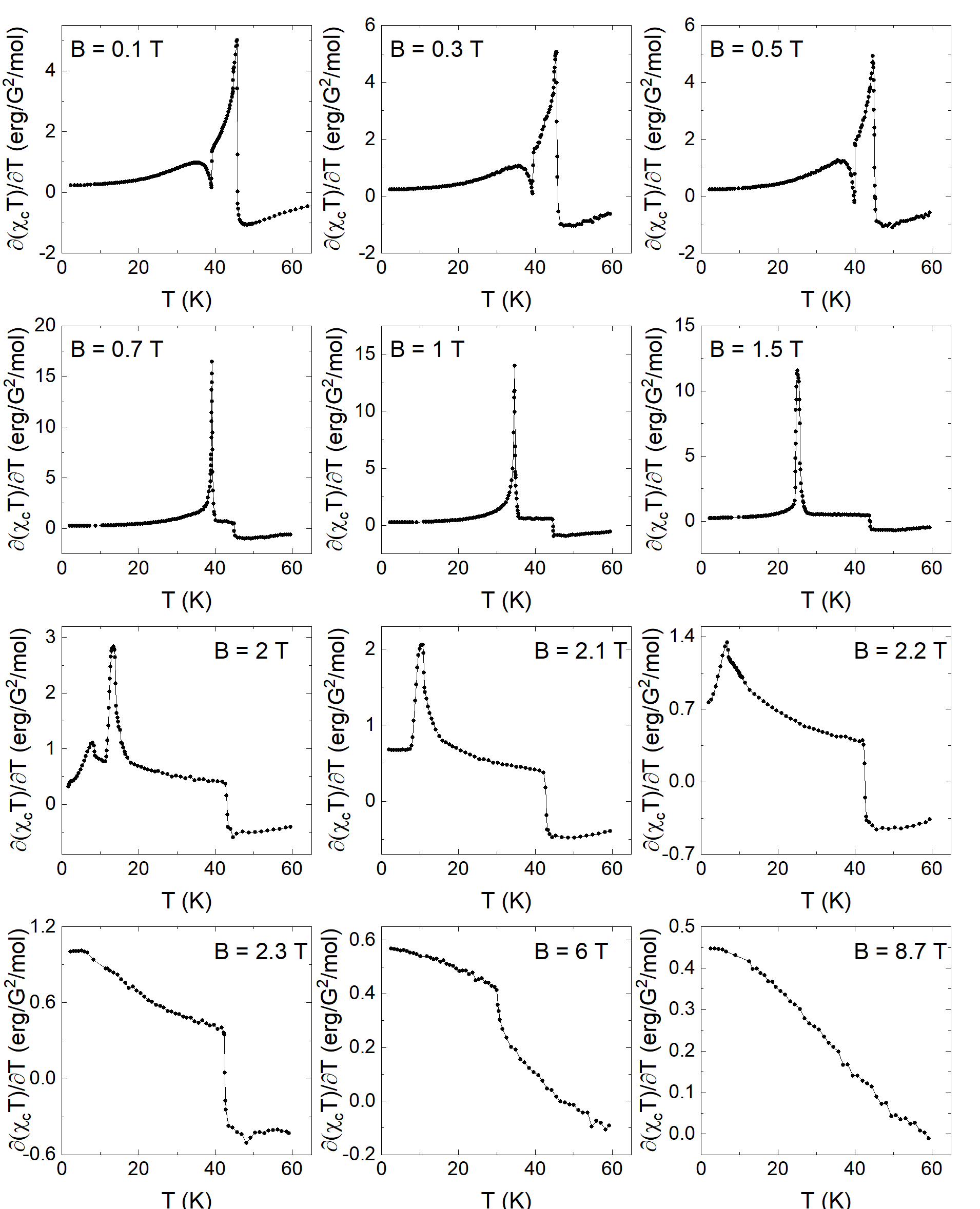}
    \caption{Fisher's specific heat $\partial(\chi_{\rm c} T)/\partial T$ at various fields up to $B \parallel c = 8.7$\,T as a function of temperature $T$. }
    \label{SM:dchiTdT}
\end{figure}

\begin{figure}[h]
    \centering
    \includegraphics[width=1\columnwidth,clip]{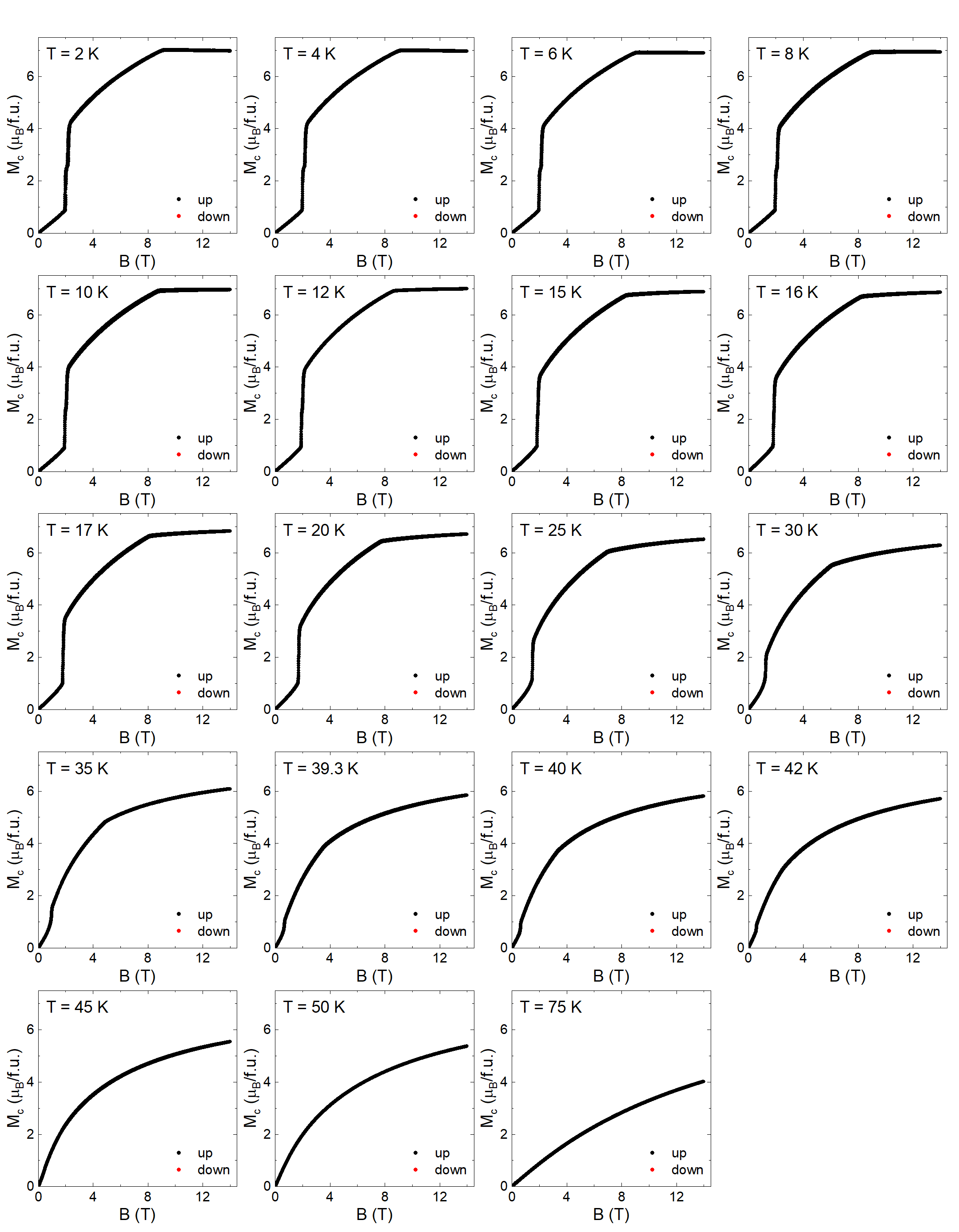}
    \caption{Isothermal magnetisation $M_{\rm c}$ at various temperatures between $T=2$\,K and $T=75$\,K as a function of the magnetic field $B \parallel c$.}
    \label{SM:MB}
\end{figure}

\begin{figure}[h]
    \centering
    \includegraphics[width=1\columnwidth,clip]{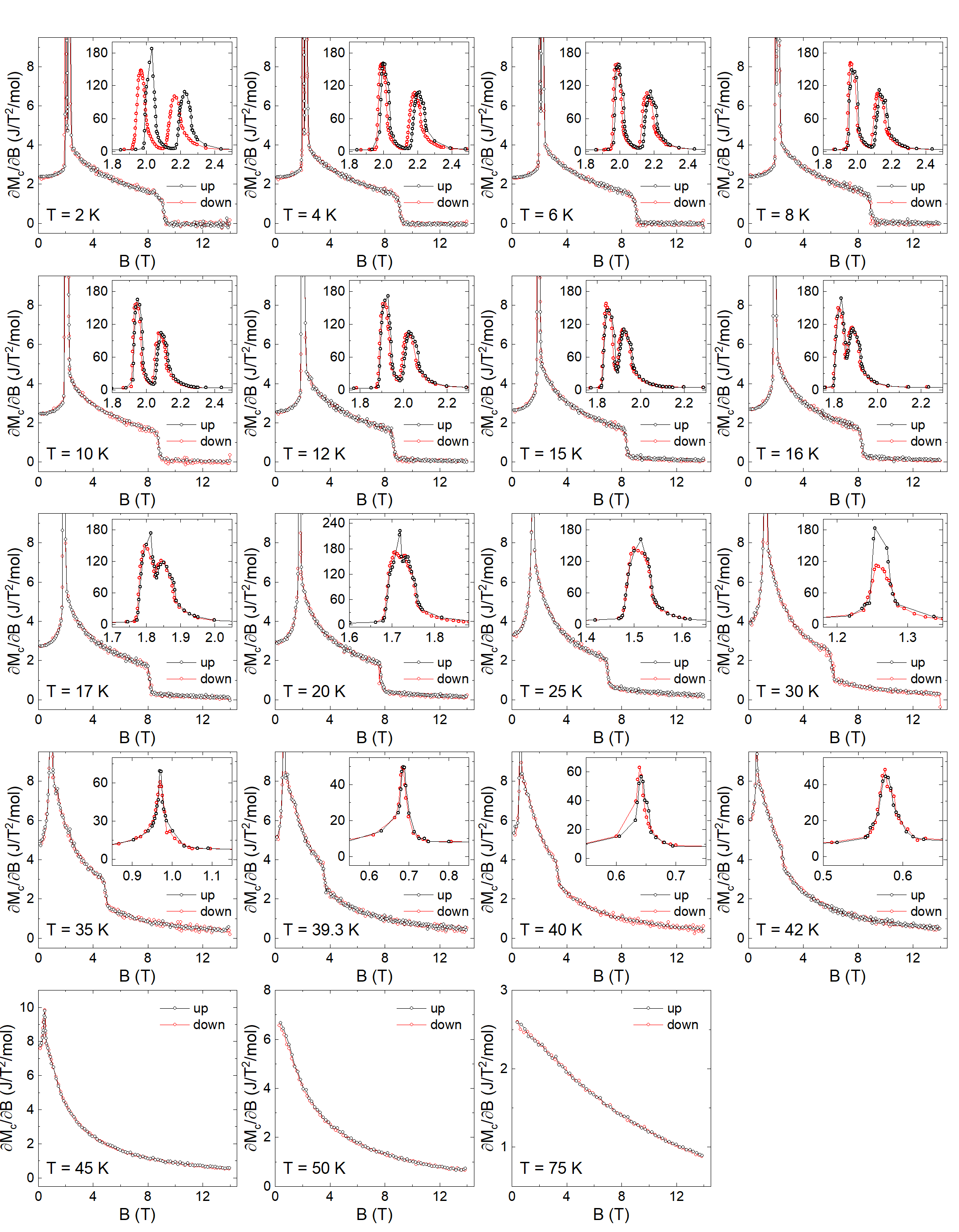}
    \caption{Magnetic susceptibility $ \partial M_{\rm c}/\partial B$ at various temperatures between $T=2$\,K and $T=75$\,K as a function of the magnetic field $B \parallel c$. The insets show the region around the low field transitions in more detail.}
    \label{SM:dMdB}
\end{figure}

\begin{figure}[h]
    \centering
    \includegraphics[width=1\columnwidth,clip]{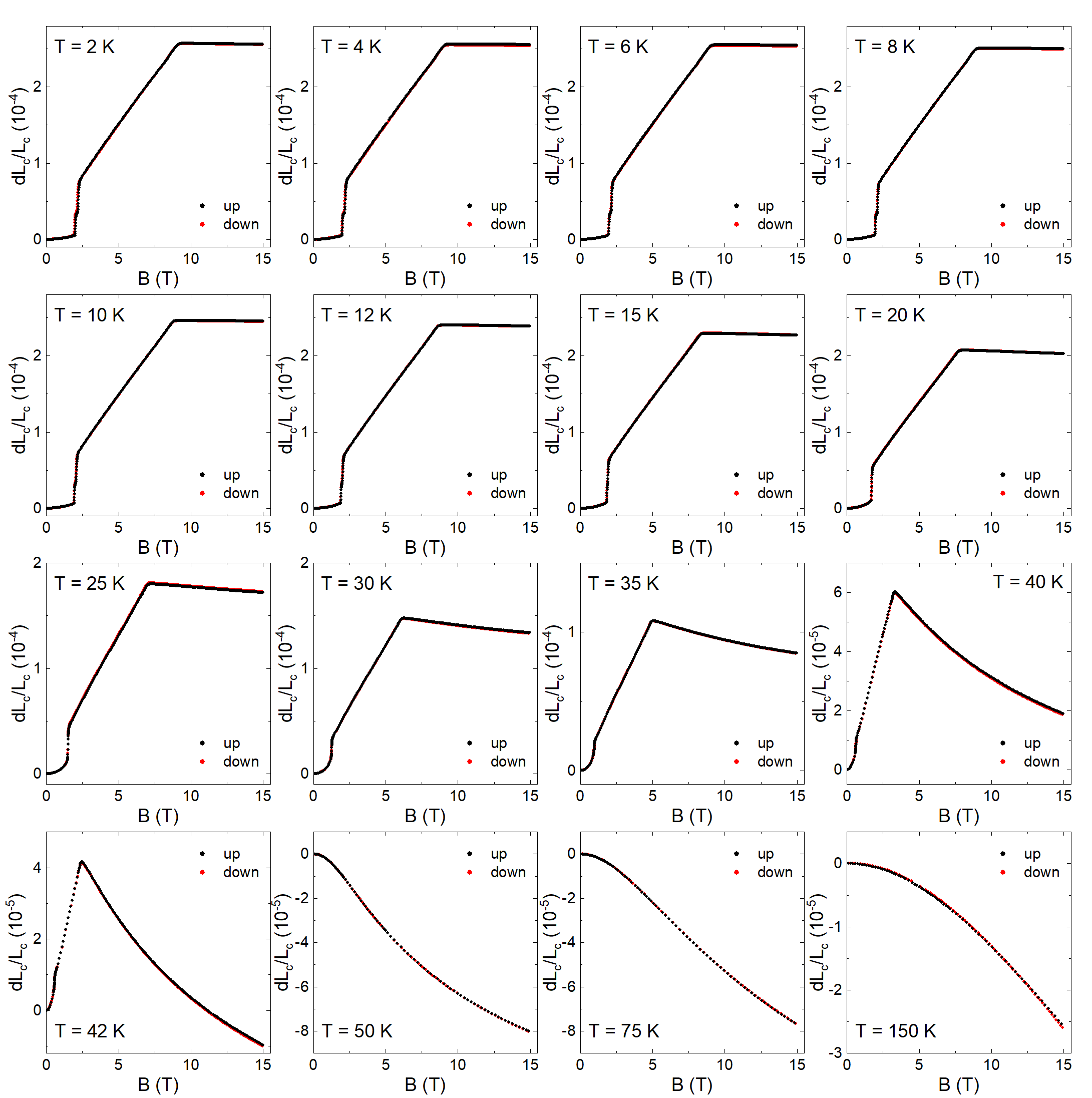}
    \caption{Magnetostriciton $dL_{\rm c}(B)/L_{\rm c}$ at various temperatures between $T=2$\,K and $T=150$\,K as a function of the magnetic field $B \parallel c$.}
    \label{SM:MS}
\end{figure}

\begin{figure}[h]
    \centering
    \includegraphics[width=1\columnwidth,clip]{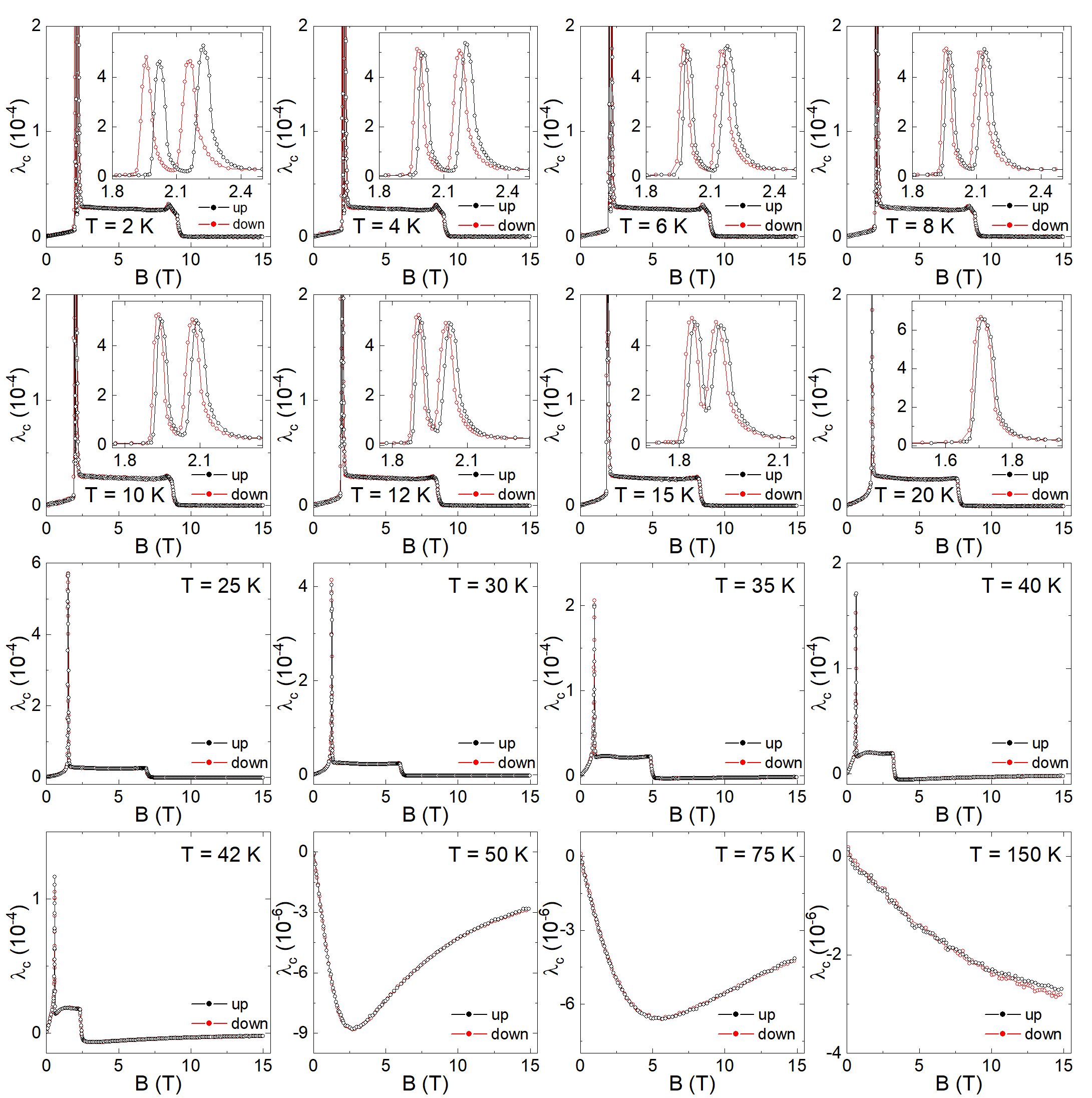}
    \caption{Magnetostriction coefficient $\lambda_{\rm c}$ at various temperatures between $T=2$\,K and $T=75$\,K as a function of the magnetic field $B \parallel c$. The insets show the region around the low field transitions in more detail.}
    \label{SM:lambda}
\end{figure}

\begin{figure}[htb]
    \centering
    \includegraphics[width=0.7\columnwidth,clip]{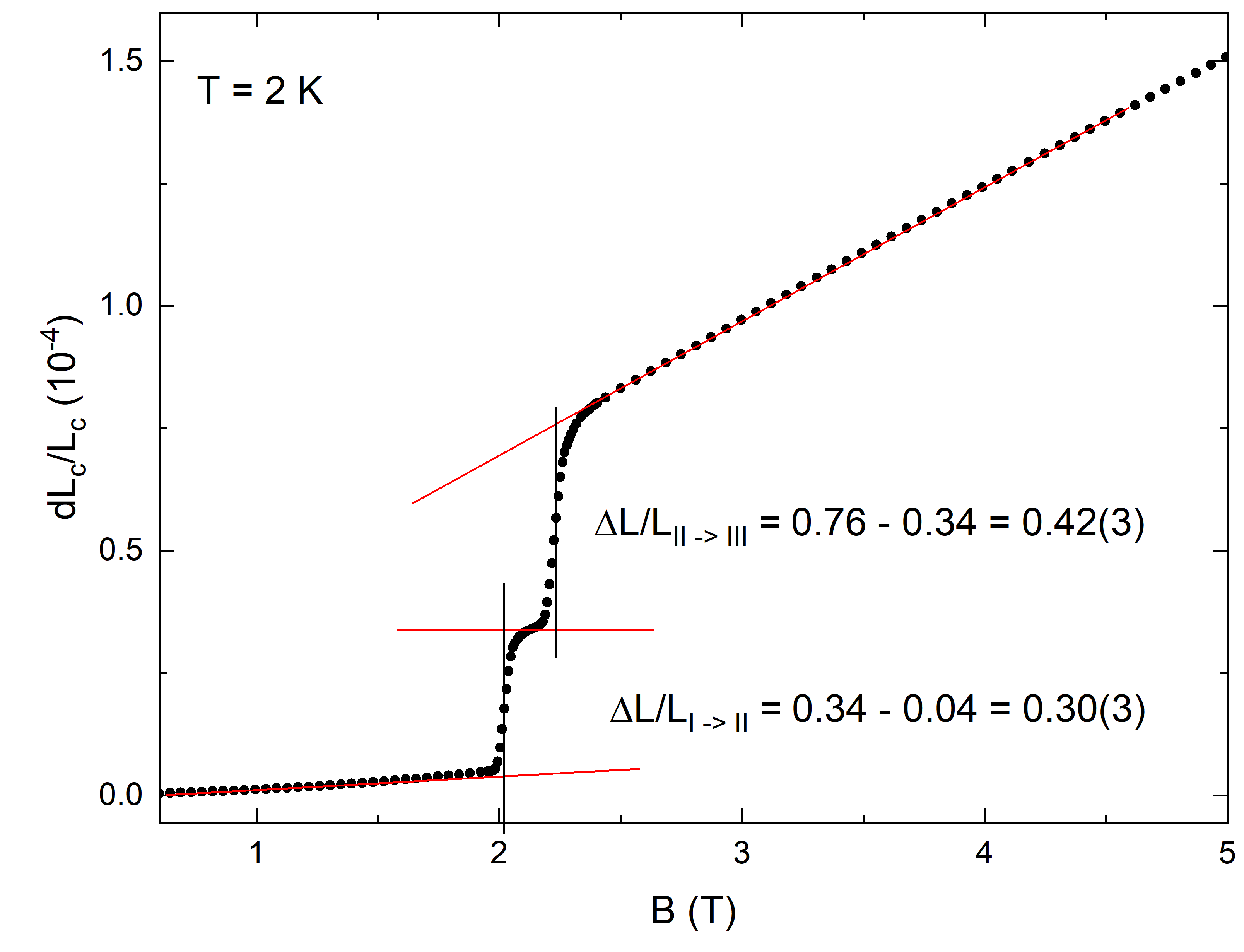}
    \caption{Determination of the jumps in magnetostriction $\Delta L_{\rm c}/L_{\rm c}$ vs. magnetic field $B$ at the phase transitions into the SKL phase (I - II) and out of the SKL phase (II - III) at T = 2\,K. In order to correctly extract the jump at the respective phase transition intervals below and above the phase transitions are approximated by a linear fit (red lines) and extrapolated to the critical field.}
    \label{SM:jump_MS_2K}
\end{figure}

\end{document}